\documentclass[10pt,letterpaper]{article}
\usepackage[top=0.85in,left=2.75in,footskip=0.75in]{geometry}

\usepackage{amsmath,amssymb}
\usepackage{changepage}
\usepackage[utf8x]{inputenc}
\usepackage{textcomp,marvosym}
\usepackage{cite}
\usepackage{nameref,hyperref}
\usepackage[right]{lineno}
\usepackage{microtype}
\DisableLigatures[f]{encoding = *, family = * }
\usepackage[table]{xcolor}
\usepackage{array}

\newcolumntype{+}{!{\vrule width 2pt}}

\newlength\savedwidth

\raggedright
\usepackage[aboveskip=1pt,labelfont=bf,labelsep=period,justification=raggedright,singlelinecheck=off]{caption}

\makeatletter
\renewcommand{\@biblabel}[1]{\quad#1.}
\makeatother

\date{}

\usepackage{lastpage,fancyhdr,graphicx}
\usepackage{epstopdf}
\fancyheadoffset[L]{2.25in}
\fancyfootoffset[L]{2.25in}
\usepackage{balance}
\usepackage{ctable}
\usepackage{multirow}
\usepackage{graphicx}
\usepackage{setspace}
\graphicspath{{../figure/}}

\usepackage{xspace}
\usepackage{verbatim}
\usepackage{latexsym}
\usepackage{subfigure}
\usepackage{bm}
\usepackage{psfrag}
\usepackage{algorithm}
\usepackage{algorithmicx}
\usepackage{algpseudocode}

\usepackage{eucal}

\usepackage[all]{xy}
\CompileMatrices

\newcommand{\mcX}{\mathcal{X}}

\newcommand{\mcZ}{\mathcal{Z}}

\newcommand{\be}{\begin{eqnarray}}
\newcommand{\ee}{\end{eqnarray}}

\DeclareMathOperator*{\argmin}{\mathrm{argmin}}

\newcommand{\intset}[1]{\cbr{1..n}}

\newcommand{\eg}{{\it e.g.}}
\newcommand{\ie}{{\it i.e.}}
\newcommand{\etal}{{\it et al.}}
\newcommand{\mylinespacing}{1}

\newcommand{\acl}{DUDE-Seq}
\newcommand{\bhlcolor}{black}
\newcommand{\tsmcolor}{black}
\newcommand{\srycolor}{black}
\newcommand{\tsmrevisioncolor}{black}
\newcommand{\tsmcolorfinal}{black}
\newcommand{\sryoonrevisioncolor}{black}
\newcommand{\srycolorfinal}{black}
\newcommand{\proofcolor}{black}

\newcommand{\bhlploscolor}{black} 
\newcommand{\tsmploscolor}{black} 
\newcommand{\sryploscolor}{black} 
\newcommand{\editageploscolor}{black}
\newcommand{\editagecolor}{black}

\newcommand{\bhlplosrevisioncolor}{black}

\begin{document}
\vspace*{0.2in}

\begin{flushleft}
{\Large
\textbf\newline{\acl: Fast, Flexible, and Robust Denoising for Targeted Amplicon Sequencing}
}
\newline
\\
Byunghan Lee\textsuperscript{1},
Taesup Moon\textsuperscript{2*},
Sungroh Yoon\textsuperscript{1,3,4*},
Tsachy Weissman\textsuperscript{5}
\\
\bigskip
\textbf{\textsuperscript{1}} Electrical and Computer Engineering, Seoul National University, Seoul, Korea
\\
\textbf{\textsuperscript{2}} College of Information and Communication Engineering, Sungkyunkwan University, Suwon, Korea
\\
\textbf{\textsuperscript{3}} Interdisciplinary Program in Bioinformatics, Seoul National University, Seoul, Korea
\\
\textbf{\textsuperscript{4}} Neurology and Neurological Sciences, Stanford University, Stanford, California, United States of America
\\
\textbf{\textsuperscript{5}} Electrical Engineering, Stanford University, Stanford, California, United States of America
\\
\bigskip
* tsmoon@skku.edu (TM), sryoon@snu.ac.kr (SY)
\end{flushleft}

\section*{Abstract}
We consider the correction of errors from nucleotide sequences produced by next-generation targeted amplicon sequencing. \textcolor{\bhlplosrevisioncolor}{The next-generation sequencing (NGS) platforms can provide a great deal of sequencing data thanks to their high throughput, but the associated error rates often tend to be high.} Denoising in high-throughput sequencing \textcolor{\editagecolor}{has} thus \textcolor{\editagecolor}{become} a crucial \textcolor{\editagecolor}{process} for boosting the reliability of downstream analyses.
Our methodology, named \acl, is derived from a general setting of reconstructing finite-valued source data corrupted by a discrete memoryless channel and \textcolor{\editagecolor}{effectively corrects} substitution and homopolymer indel errors, the two major types of sequencing errors in most high-throughput targeted amplicon sequencing platforms. Our experimental studies with real and simulated datasets suggest that the proposed \acl~not only outperforms existing alternatives in terms of error-correction \textcolor{\editagecolor}{capability} and time efficiency, but also \textcolor{\editagecolor}{boosts} the reliability of downstream analyses. Further, the flexibility of \acl~enables \textcolor{\editagecolor}{its robust application} to different sequencing platforms and analysis pipelines \textcolor{\editagecolor}{by} simple update\textcolor{\editagecolor}{s} of the noise model.
\acl~is available at \href{http://data.snu.ac.kr/pub/dude-seq}{http://data.snu.ac.kr/pub/dude-seq}.

\section*{Author Summary}
\textcolor{\srycolor}{
Next-generation sequencing (NGS) has already become a fundamental means for understanding a variety of biological processes in living organisms, creating numerous academic and practical opportunities. The success of NGS can largely be accredited to the low cost and high throughput of mainstream NGS technology, \textcolor{\proofcolor}{but that} inevitably incur\textcolor{\proofcolor}{s} a sacrifice in robustness measured in terms of error rates in sequenced reads. Denoising in NGS is thus a crucial component in many NGS analysis pipelines in order to ensure the reliability and quality of sequencing results. In this paper, we propose a new denoising algorithm named \acl, which possesses flavors connected to existing denoising methodologies such as $k$-mer based, multiple sequence alignment-based, and statistical error model-based techniques, \textcolor{\proofcolor}{to} effectively overcoming their limitations.
\textcolor{\bhlplosrevisioncolor}{As the sequencing coverage becomes deeper, context-counting vectors can accumulate more probable contexts, and the robustness of denoising normally improves, hence, we focus on the targeted amplicon sequencing.}
Our thorough evaluation efforts lead us to \textcolor{\proofcolor}{conclude} that the proposed \acl~algorithm is effective in removing substitution errors and homopolymer errors \textcolor{\proofcolor}{that} frequently \textcolor{\bhlploscolor}{occur} in applications of NGS \textcolor{\proofcolor}{for} targeted amplicon sequencing. We also anticipate that the flexibility of \acl~will make it a versatile building block of other NGS pipelines that need efficiency and robustness \textcolor{\proofcolor}{for} large-scale sequence processing, such as \textcolor{\proofcolor}{the} denoising workflow involved in the emerging nanopore sequencing technology.}

\section*{Introduction}
A new generation of high-throughput, low-cost sequencing technologies, referred to as \emph{next-generation sequencing} (NGS) \textcolor{\editagecolor}{technologies}~\cite{metzker2010sequencing}, is reshaping biomedical research\textcolor{\editagecolor}{,} including large-scale comparative and evolutionary studies~\cite{astbury1961molecular,bateson1894materials,riesenfeld2004metagenomics}. Compared with automated Sanger sequencing, NGS \textcolor{\editagecolor}{platforms} produce significantly shorter reads in large quantit\textcolor{\proofcolor}{ies}, posing various new computational challenges~\cite{pop2008bioinformatics}.

\textcolor{\editagecolor}{There are} \textcolor{\bhlplosrevisioncolor}{several} DNA sequencing methodologies \textcolor{\proofcolor}{that use} NGS~\textcolor{\bhlploscolor}{\cite{shendure2008next,goodwin2016coming}} \textcolor{\bhlplosrevisioncolor}{including} whole genome sequencing (WGS), chromatin immunoprecipitation (ChIP) sequencing, and targeted sequencing. WGS is \textcolor{\editagecolor}{used to} analyze the genome of an organism to \textcolor{\bhlcolor}{capture all variants and} identify potential causative variants; \textcolor{\bhlploscolor}{\textcolor{\editageploscolor}{it} is also used for \textit{de novo} genome assembly.} ChIP sequencing identif\textcolor{\proofcolor}{ies} genome-wide DNA binding sites for transcription factors and other proteins.
Targeted sequencing (\eg, exome sequencing and amplicon sequencing), the focus of this paper, is a cost-effective method that enables researchers to focus on investigating areas of interest that are likely to be involved in \textcolor{\editagecolor}{a particular} phenotype.
According to \textcolor{\editagecolor}{previous studies}~\cite{bamshad2011exome,jamuar2015clinical}, \textcolor{\bhlploscolor}{targeted sequencing often \textcolor{\editageploscolor}{results in the} complete coverage of exons of disease\textcolor{\editageploscolor}{-}related genes\textcolor{\editageploscolor}{,} while alternative \textcolor{\editageploscolor}{methods result in} approximately 90--95\% \textcolor{\editageploscolor}{coverage}. Hence, in clinical setting\textcolor{\editageploscolor}{s}, researchers tend to rely on targeted sequencing for diagnostic evaluation\textcolor{\editageploscolor}{s}.}

To detect sequences \textcolor{\editagecolor}{based on} fluorescent labels at the molecular level, NGS technologies normally rely on imaging systems requir\textcolor{\editagecolor}{ing} templates \textcolor{\editagecolor}{that are amplified by} emulsion polymerase chain reaction (PCR) or solid-phase amplification~\cite{metzker2010sequencing}. These amplification and imaging processes can \textcolor{\editagecolor}{generate} erroneous reads, the origin of which can be traced \textcolor{\proofcolor}{to the} incorrect determination of homopolymer lengths, \textcolor{\editagecolor}{the} erroneous insertion/deletion/substitution of nucleotide bases, and PCR chimera\textcolor{\editagecolor}{s}~\cite{shendure2008next}. Substitution errors dominate \textcolor{\editagecolor}{for} many platforms\textcolor{\editagecolor}{,} including Illumina, while homopolymer errors\textcolor{\editagecolor}{,} manifested as insertions and deletions (indels)\textcolor{\editagecolor}{,} are also abundant \textcolor{\editagecolor}{for} 454 pyrosequencing and Ion Torrent.

Erroneous reads must be properly handled \textcolor{\proofcolor}{because} they complicate downstream analyses (\eg, variant calling and genome assembly), often lowering the quality of the whole analysis pipeline\textcolor{\bhlploscolor}{~\cite{goodwin2016coming}.} Soft clipping, \textcolor{\editagecolor}{in which} 3'-ends of a read \textcolor{\editagecolor}{are trimmed} based on the quality scores of individual bases, may be the simplest approach, but it results in \textcolor{\editagecolor}{a} loss of information~\cite{yang2013survey}. More sophisticated methods \textcolor{\proofcolor}{focus on} detecting and correcting errors in sequence data~\cite{ilie2011hitec,kao2011echo,kelley2010quake,qu2009efficient,salmela2010correction,salmela2011correcting,schroder2009shrec,wijaya2009recount,yang2011repeat,yang2010reptile}. Given the widespread use of Illumina sequencing platforms, most error-correction algorithms have targeted substitution errors~\cite{yang2013survey}.

As summarized in recent \textcolor{\editagecolor}{reviews}~\cite{yang2013survey,laehnemann2015denoising}, current error-correction methods for NGS \textcolor{\editagecolor}{data} can be categorized as follows: $k$-mer (\ie, oligonucleotide\textcolor{\editagecolor}{s} of length $k$) frequency/spectrum\textcolor{\editagecolor}{-}based, multiple sequence alignment (MSA)\textcolor{\editagecolor}{-}based, and statistical error model\textcolor{\proofcolor}{-}based methods.
The idea \textcolor{\proofcolor}{behind} $k$-mer\textcolor{\editagecolor}{-}based methods~\cite{kelley2010quake,yang2010reptile,medvedev2011error,nikolenko2013bayeshammer,greenfield2014blue,lim2014trowel} is \textcolor{\srycolor}{to create a list of \textcolor{\proofcolor}{``}trusted\textcolor{\proofcolor}{''} $k$-mers from the input reads and correct untrusted $k$-mers \textcolor{\editagecolor}{based on} a consensus represented by this spectrum. In addition to the length of \textcolor{\editagecolor}{the} $k$-mer, coverage ($k$-mer occurrences) information is important to determine trusted $k$-mers.} Under the assumption that errors are rare and random and that coverage is uniform, for sufficiently large $k$, it is reasonable to expect that \textcolor{\bhlplosrevisioncolor}{most errors} alter $k$-mers to inexistent ones in a genome. Thus, \textcolor{\editagecolor}{for} high-coverage genome sequences \textcolor{\editagecolor}{obtained by} NGS, we may identify suspicious $k$-mers and correct them \textcolor{\editagecolor}{based on} a consensus.
MSA\textcolor{\proofcolor}{-}based methods~\cite{salmela2011correcting,kao2011echo,bragg2012fast} work by aligning \textcolor{\srycolor}{related sequences according to their similarities and \textcolor{\editagecolor}{correcting} aligned reads\textcolor{\editagecolor}{,} usually \textcolor{\editagecolor}{based on} a consensus in an alignment column\textcolor{\editagecolor}{,} using various techniques. This alignment-based scheme is inherently well-suited for correcting indel errors. Early methods suffered from computational issues, but recent approaches utilize advanced indexing techniques to expedite the alignments.}
\textcolor{\editagecolor}{In} statistical error model\textcolor{\proofcolor}{-}based methods~\cite{meacham2011identification,yin2013premier,schulz2014fiona}\textcolor{\editagecolor}{,}
\textcolor{\srycolor}{a statistical model \textcolor{\editagecolor}{is developed} to capture the sequencing process\textcolor{\editagecolor}{,} including error generation. In this regard, an empirical confusion model \textcolor{\editagecolor}{is often created} from datasets, exploiting the information obtained from, \eg, alignment results\textcolor{\editagecolor}{,} Phred quality scores (a measure of the quality of nucleobases generated by automated DNA sequencing)~\cite{ewing1998base}\textcolor{\editagecolor}{, or other parameters}.}

\textcolor{\tsmrevisioncolor}{While the above methods often \textcolor{\editagecolor}{exhibit good} performance \textcolor{\editagecolor}{for} various platforms, they also have several limitations. First, $k$-mer\textcolor{\editagecolor}{-}based schemes tend to be \textcolor{\bhlplosrevisioncolor}{ineligible} when the coverage is expected to vary over the queried sequences, as in transcriptomics, metagenomics, heterogen\textcolor{\editagecolor}{e}ous cell samples\textcolor{\editagecolor}{,} or pre-amplified libraries \cite{laehnemann2015denoising}.  Second, MSA\textcolor{\proofcolor}{-}based methods, \textcolor{\editagecolor}{which do} not suffer from \textcolor{\proofcolor}{the} above \textcolor{\editagecolor}{issue related to} non-uniform coverage, often \textcolor{\editagecolor}{require the application of} heuristic and sophisticated consensus decision rules for the aligned columns, and such rules may be sensitive to specific applications or sequencing platforms.
Third, statistical error model\textcolor{\proofcolor}{-}based methods typically \textcolor{\editagecolor}{use} computationally expensive schemes (e.g., expectation-maximization) \textcolor{\editagecolor}{owing} to additional stochastic modeling assumptions \textcolor{\editagecolor}{for} the underlying DNA sequences. Moreover, little attention is given to the validity and accuracy of such modeling assumptions, let alone to theoretical analysis \textcolor{\editagecolor}{of} whether near optimum or sound error-correction performance is attained. Finally, many existing schemes applying the three methods often return only representative (consensus) denoised sequences created by merging input sequences\textcolor{\proofcolor}{;} hence, the number of sequences is often not preserved after denoising. In some applications, this may result in inconsistencies in downstream analyses.
\textcolor{\editagecolor}{To address} these limitations, many existing tools combine the three methods in a complementary \textcolor{\editagecolor}{manner to improve} performance~\textcolor{\bhlploscolor}{\cite{yang2013survey,laehnemann2015denoising}.}}

\textcolor{\tsmploscolor}{In this paper, as an alternative, we \textcolor{\editageploscolor}{applied} an algorithm called Discrete Universal DEnoiser (DUDE)~\cite{weissman2005universal} \textcolor{\editageploscolor}{for accurate} DNA sequence denoising. DUDE was developed for a general setting of reconstructing sequences with finite-valued components (source symbols) corrupted by a
noise mechanism that corrupts each source symbol independently and statistically identically. \textcolor{\tsmploscolor}{In \textcolor{\editageploscolor}{the} DNA denoising literature, such \textcolor{\editageploscolor}{a} noise model is equivalent to the confusion matrix commonly used in statistical error-model\textcolor{\editageploscolor}{-}based methods.} \textcolor{\editageploscolor}{As demonstrated in the} original paper~\cite{weissman2005universal}\textcolor{\editageploscolor}{, DUDE exhibits} rigorous performance guarantee for the following setting;
even when no stochastic modeling assumptions are made \textcolor{\editageploscolor}{for} the underlying clean source data, only with the assumption of \emph{known} noise mechanism,
DUDE is shown to universally attain the optimum denoising performance
for \emph{any} source data \textcolor{\editageploscolor}{the data increase}. We note that the above setting of DUDE naturally fits the setting \textcolor{\editageploscolor}{for} DNA sequence denoising\textcolor{\editageploscolor}{,} \ie, it is difficult to \textcolor{\editageploscolor}{establish} accurate stochastic models for clean DNA sequences, but it is simple and fairly realistic to assume noise models (\ie, confusion matrices) for sequencing devices \textcolor{\editageploscolor}{based on} reference sequences.}

\textcolor{\tsmrevisioncolor}{The DUDE algorithm, which will be explained in details \textcolor{\srycolor}{in the next section}, possesses flavors that are somewhat connected to all three representative methods mentioned above, in a single scheme. \textcolor{\editagecolor}{Specifically}, DUDE works with double-sided contexts of \textcolor{\proofcolor}{a} fixed size \textcolor{\proofcolor}{that} are analogous to $k$-mers. Moreover, like MSA, DUDE applies a denoising decision rule \textcolor{\editagecolor}{to} each noisy symbol based on aggregated information over certain positions in the reads. \textcolor{\proofcolor}{However}, unlike MSA\textcolor{\proofcolor}{,} which makes a decision based on the information collected from the symbols in the same aligned column, DUDE makes a decision using the information collected from positions with the same double-sided context. Finally, the denoising decision rule of DUDE utilizes information \textcolor{\proofcolor}{from} \textcolor{\tsmploscolor}{the assumed noise model}\textcolor{\editagecolor}{,} like \textcolor{\editagecolor}{in} most statistical error model\textcolor{\proofcolor}{-}based methods, but does not assume any stochastic model on the underlying sequence, thus result\textcolor{\proofcolor}{ing} in a computationally efficient method. The \textcolor{\proofcolor}{method} of incorporating the \textcolor{\tsmploscolor}{noise} model is also simple, mak\textcolor{\proofcolor}{ing} it easy to flexibly apply DUDE to different sequencing platforms by \textcolor{\editagecolor}{simply} changing the \textcolor{\tsmploscolor}{confusion matrix model} in the algorithm.}

\textcolor{\tsmrevisioncolor}{With the above unique nature of \textcolor{\proofcolor}{the} DUDE algorithm, we show in our experiments that it outperforms other state-of-the-art schemes\textcolor{\editagecolor}{,} particularly for \textcolor{\editagecolor}{applications to} targeted amplicon sequencing.}
Specifically, among the applicable areas of targeted amplicon sequencing (\eg, cancer gene, 16S rRNA, plant, and animal sequencing~\cite{schirmer2015insight}), we used 16S rRNA benchmark datasets \textcolor{\editagecolor}{obtained with} different library preparation methods and DNA polymerases to confirm the robustness of our algorithm \textcolor{\editagecolor}{across various} sequencing preparation methods.
\textcolor{\bhlploscolor}{Targeted amplicon sequencing datasets often have deeper sequencing coverage than \textcolor{\editageploscolor}{those of} WGS or ChIP datasets, which frequently makes conventional $k$-mer-based techniques \textcolor{\editageploscolor}{often} suffer from the amplification bias problem~\cite{yan2016coverage}. By contrast, \textcolor{\editageploscolor}{for} \acl, as the sequencing coverage becomes deeper, context-counting vectors can accumulate more probable contexts, and the robustness of denoising \textcolor{\editageploscolor}{typically} improves.}
We apply two versions of DUDE separately for substitution and homopolymer errors, the two major types of sequencing error. For substitution error\textcolor{\tsmrevisioncolor}{s}, our approach \textcolor{\editagecolor}{directly} utilizes the original DUDE with appropriate adaptation to DNA sequences and is applicable to reads \textcolor{\editagecolor}{generated by} any sequencing platform. For homopolymer error\textcolor{\tsmrevisioncolor}{s}, however, we do not apply the original DUDE\textcolor{\editagecolor}{,} which was developed in a framework that does not cover errors of the homopolymer type. To correct homopolymer errors, we \textcolor{\proofcolor}{therefore} adopt a variant of DUDE for general-output channels \cite{dembo2005universal}. Our homopolymer-error correction is applicable to cases in which base-called sequences and the underlying flowgram intensities are available (\eg, pyrosequencing and Ion Torrent). For brevity, we refer to both of \textcolor{\editagecolor}{these} DUDE-based approaches as \acl\textcolor{\editagecolor}{, but the correction type will be easily distinguishable by the reader.}

\section*{Discrete Universal DEnoiser (DUDE)}\label{sec:related_work}
\begin{figure}[!h]
	\centering
	\includegraphics[width=\linewidth]{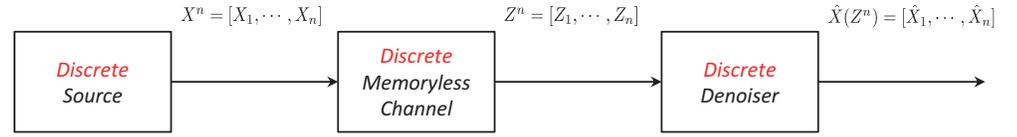}
	\caption{{\bf The general setting of discrete denoising.}}
	\label{fig:general-setting}
\end{figure}

\textcolor{\tsmrevisioncolor}{In this section, we formally introduce \textcolor{\proofcolor}{the} DUDE algorithm \textcolor{\proofcolor}{along} with \textcolor{\editagecolor}{its} notation and its connection to DNA sequence denoising.} Fig~\ref{fig:general-setting} shows the concrete setting of the discrete denoising problem.
We denote the underlying source data as $\{x_i\}$ and assume each component takes values in some finite set $\mcX$. The resulting noisy version of the source corrupted by \textcolor{\tsmploscolor}{a noise mechanism} is denoted as $\{Z_i\}$, and its components take values in, again, some finite set $\mcZ$. \textcolor{\tsmploscolor}{As mentioned in \textcolor{\editageploscolor}{the} Introduction, DUDE assumes that the noise mechanism injects noise\textcolor{\editageploscolor}{s} that are independent and statistically identical, and such \textcolor{\editageploscolor}{a} mechanism is often referred to as \textcolor{\editageploscolor}{a} Discrete Memoryless Channel (DMC) in information theory.}
The DMC is completely characterized by the channel transition matrix\textcolor{\tsmploscolor}{, also known as the confusion matrix,} $\mathbf\Pi\in\mathbb{R}^{|\mcX|\times|\mcZ|}$, of which the $(x,z)$-th element, $\Pi(x,z)$, stands for $\text{Pr}(Z_i=z|x_i=x)$, \ie, the conditional probability \textcolor{\editagecolor}{that} the noisy symbol tak\textcolor{\editagecolor}{es} value $z$\textcolor{\editagecolor}{,} given \textcolor{\editagecolor}{that} the original source symbol \textcolor{\editagecolor}{is} $x$. \textcolor{\tsmploscolor}{We denote random variables with uppercase letters and the individual samples of random variables or deterministic symbols with lowercase letters. Thus, the underlying source data, which \textcolor{\editageploscolor}{are} treated by DUDE as individual sequence\textcolor{\editageploscolor}{s} (and not a stochastic process), \textcolor{\editageploscolor}{are} denoted \textcolor{\editageploscolor}{by} the lowercase $\{x_i\}$, and the noise-corrupted sequence\textcolor{\editageploscolor}{s}, \textcolor{\editageploscolor}{\ie,} sequence\textcolor{\editageploscolor}{s} of random variables, \textcolor{\editageploscolor}{are} denoted \textcolor{\editageploscolor}{by} uppercase $\{Z_i\}$.}
Furthermore, throughout this paper, we generally denote a sequence ($n$-tuple) as $a^n=(a_1,\ldots,a_n)$, \textcolor{\editagecolor}{for example, where} $a_i^j$ refers to the subsequence $(a_i,\ldots,a_j)$.

As shown in Fig~\ref{fig:general-setting}, a discrete denoiser observes the entire noisy data $Z^n$ and reconstructs the original data with $\hat{X}^n=(\hat{X}_1(Z^n),\ldots,\hat{X}_n(Z^n))$. The goodness of the reconstruction by a discrete denoiser $\hat{X}^n$ is measured by the average loss,
\begin{equation}
L_{\hat{X}^n}(x^n,Z^n) = \frac{1}{n}\sum_{i=1}^n\Lambda(x_i,\hat{X}_i(Z^n)),\label{eq:avg_loss}
\end{equation}
where $\Lambda(x_i,\hat{x}_i)$ is a single-letter loss function that measures the loss incurred by estimating $x_i$ with $\hat{x}_i$ at location $i$. The loss function can be also represented with a loss matrix $\mathbf{\Lambda}\in\mathbb{R}^{|\mcX|\times|\hat{\mcX}|}$.

DUDE in \cite{weissman2005universal} is a two-pass algorithm that has linear complexity \textcolor{\editagecolor}{with respect to} the data size $n$. During the first pass, given the realization of the noisy sequence $z^n$, the algorithm collects the statistics vector
\be
\mathbf{m}(z^n, l^k,r^k)[a] = \big|\{i: k+1\leq i \leq n-k, z_{i-k}^{i+k} = l^kar^k\}\big|, \nonumber
\ee
for all $a\in\mcZ$, which is the count of the occurrence of the symbol $a\in\mcZ$ along the noisy sequence $z^n$ that has the \emph{double-sided context} $(l^k, r^k)\in\mcZ^{2k}$. \textcolor{\tsmrevisioncolor}{Note \textcolor{\editagecolor}{that} $\mathbf{m}$ is similar to the counts across the aligned columns for the simple majority voting in MSA-based denoising methods. \textcolor{\editagecolor}{However,} in DUDE, the count is collected regardless of \textcolor{\proofcolor}{whether} the positions in the reads \textcolor{\proofcolor}{are} aligned or not, but \textcolor{\editagecolor}{considering} whether the position has the same context. \textcolor{\editagecolor}{Additionally}, the context length $k$ is analogous to the $k$-mer length.} Once the $\mathbf{m}$ vector is collected, for the second pass, DUDE then applies the rule
\be
\hat{X}_i(z^n) =\arg\min_{\hat{x}\in\mcX}\mathbf{m}^T(z^n,z_{i-k}^{i-1},z_{i+1}^{i+k})\mathbf{\Pi}^{-1}[\lambda_{\hat{x}}\odot \pi_{z_i}]\label{eq:dude_rule}
\ee
for each $k+1\leq i\leq n-k$, where $\pi_{z_i}$ is the $z_i$-th column of the channel matrix $\mathbf{\Pi}$, and $\lambda_{\hat{x}}$ is the $\hat{x}$-th column of the loss matrix $\mathbf{\Lambda}$. \textcolor{\tsmrevisioncolor}{Furthermore, $\odot$ stands for the element-wise product operator for two vectors.} \textcolor{\tsmploscolor}{The \textcolor{\editageploscolor}{intuitive explanation of} (\ref{eq:dude_rule}) is \textcolor{\editageploscolor}{as} follow\textcolor{\editageploscolor}{s}: when we rearrange the right-hand side of (\ref{eq:dude_rule}), we obtain
\be
(\ref{eq:dude_rule})&=&\arg\min_{\hat{x}\in\mcX}\lambda_{\hat{x}}^T\big \{\pi_{z_i}\odot\mathbf{\Pi}^{-T}\mathbf{m}^T(z^n,z_{i-k}^{i-1},z_{i+1}^{i+k})\big\}\label{eq:dude_rule_2},
\ee
and we can show that $\pi_{a}\odot\mathbf{\Pi}^{-T}\mathbf{m}^T(z^n,l^k,r^k)$ approximates the empirical count vector of the underlying \emph{clean} symbol at the middle location that resulted in the noisy context $l^kar^k$.
Thus, the denoising rule (\ref{eq:dude_rule}), re-expressed in (\ref{eq:dude_rule_2}), \textcolor{\editageploscolor}{finds} a reconstruction symbol $\hat{x}$ that minimizes the expected loss with respect to the \emph{empirical estimate} (\textcolor{\editageploscolor}{obtained} by utilizing the inverse of $\mathbf{\Pi}$) of the count vector
of the underlying $x_i$ given the noisy context $z_{i-k}^{i+k}$. At \textcolor{\editageploscolor}{a} high level, DUDE is not a simple majority voting rule based on $\mathbf{m}$\textcolor{\editageploscolor}{; instead,} it incorporates the DMC model $\mathbf{\Pi}$ (the confusion matrix) and loss function $\mathbf{\Lambda}$ to \textcolor{\editageploscolor}{obtain} a more accurate estimation of the clean source symbol. For more detailed and rigorous arguments on the \textcolor{\editageploscolor}{intuitive description} of (\ref{eq:dude_rule}), we refer readers to the original paper~\cite[Section IV-B]{weissman2005universal}.}

\textcolor{\tsmploscolor}{Note \textcolor{\editageploscolor}{that} formula (\ref{eq:dude_rule}) assumes $\mcX=\mcZ=\hat{\mcX}$ and $\mathbf{\Pi}$ is invertible for simplicity, but \textcolor{\sryploscolor}{Weissman et al.~\cite{weissman2005universal} deal} with more general cases as well. The form of (\ref{eq:dude_rule}) \textcolor{\tsmploscolor}{also} shows that DUDE is a sliding window denoiser with window size $2k+1$\textcolor{\proofcolor}{;} \ie, DUDE returns the same denoised symbol at all locations with the same value $z_{i-k}^{i+k}$. DUDE is guaranteed \textcolor{\proofcolor}{to} attain the optimum performance by the sliding window denoisers with the same window size as the observation length $n$ increases. For more details on the theoretical performance analyses, \textcolor{\editageploscolor}{see} \textcolor{\sryploscolor}{Weissman et al.}~\cite[Section V]{weissman2005universal}.}

The original DUDE dealt exclusively with the case of $|\mcX|$ and $|\mcZ|$ finite. \textcolor{\bhlploscolor}{Dembo and Weissman~\cite{dembo2005universal} generalized} DUDE to the case of discrete input and general output channels; the noisy outputs \textcolor{\proofcolor}{do} not have \textcolor{\proofcolor}{to have} their values in some finite set, but can have continuous values as well. As in \cite{weissman2005universal}, the memoryless noisy channel model, which is characterized \textcolor{\proofcolor}{in this case} by the set of densities $\{f_{x}\}_{x\in\mcX}$, was assumed \textcolor{\editagecolor}{to be} known. As shown in \cite[Fig~1]{dembo2005universal}, the crux of the arguments is to apply a scalar quantizer $Q(\cdot)$ to each continuous-valued noisy output $\{Y_i\}$ and \textcolor{\editagecolor}{to} derive a virtual DMC, $\mathbf{\Gamma}\in\mathbb{R}^{|\mcX|\times|\mcZ|}$, between the discrete input $\{X_i\}$ and the quantized (hence, discrete) output $\{Z_i\}$. Such $\mathbf{\Gamma}$ can be readily obtained by the knowledge of $\{f_{x}\}_{x\in\mcX}$ and evaluating the following integral for each $(x,z)$: $\Gamma(x,z) = \int_{y:Q(y)=z}f_x(y)dy$.
Once the virtual DMC is obtained, the rest of the algorithm in \cite{dembo2005universal} proceeds similarly as the original DUDE; \textcolor{black}{\textcolor{\editagecolor}{specifically}, \textcolor{\proofcolor}{it} obtain\textcolor{\proofcolor}{s} the statistics vector $\mathbf{m}$ for the quantized noisy outputs $\{Z_i\}$ during the first pass, \textcolor{\proofcolor}{and} then appl\textcolor{\proofcolor}{ies} a sliding window denoising rule similar to (\ref{eq:dude_rule}), which depends on the statistics vector $\mathbf{m}$, the virtual DMC $\mathbf{\Gamma}$, $\{f_x\}_{x\in\mathcal X}$, and the noisy sequence $Y^n$, during the second pass. A concrete denoising rule can be found in \cite[Eqs. (16),(19), and (20)]{dembo2005universal}.
In \cite{dembo2005universal}, a formal analysis of the generalized DUDE shows that it attains the optimum denoising performance among sliding window denoisers with the same window size, that base their denoising decisions on the original continuous-valued outputs $Y^n$. We refer readers to the paper for more details. In the next section, we show how we adopt this generalized DUDE in our \acl~to correct homopolymer errors in DNA sequencing.}

\section*{\acl~: DUDE for DNA Sequence Denoising}
\subsection*{Substitution Errors.}
\begin{figure}[!h]
	\centering
	\includegraphics[width=0.7\linewidth]{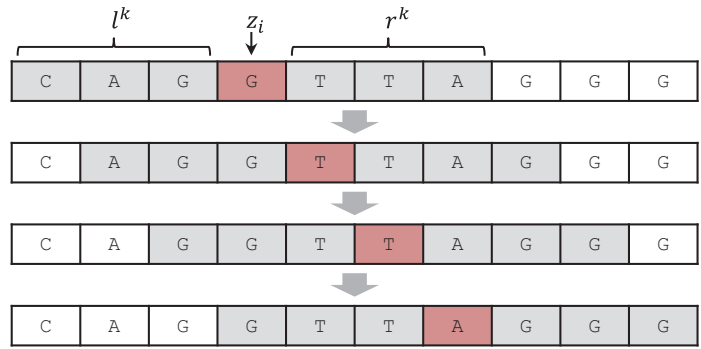}
	\begin{spacing}{\mylinespacing}
		\caption{{\bf A sliding window procedure of the DUDE-Seq \textcolor{\tsmrevisioncolor}{with} the context size $\mathbf{k = 3}$.} During the first pass, \acl~updates the $\mathbf{m}(z^{\textcolor{\tsmrevisioncolor}{n}},l^3,r^3)$ for the encountered double-sided contexts $(l^3,r^3)$. Then, for the second pass, \acl~uses the obtained $\mathbf{m}(z^{\textcolor{\tsmrevisioncolor}{n}},l^3,r^3)$ and (\ref{eq:dude_rule}) for the denoising.}
		\label{fig:context-def}
	\end{spacing}
\end{figure}

As described in the previous section, the setting of \textcolor{\proofcolor}{the} original DUDE algorithm naturally aligns with the setting of substitution-error \textcolor{\editagecolor}{correction} in DNA sequence denoising. We can set $\mcX=\mcZ= \{\texttt{A},\texttt{C},\texttt{G},\texttt{T}\}$, and the loss function as the Hamming loss, namely, $\Lambda(x,\hat{x})=0$, if $x=\hat{x}$, and $\Lambda(x,\hat{x})=1$, otherwise. Then, the two-pass sliding window procedure of DUDE \textcolor{\proofcolor}{for} collecting the statistics vector $\mathbf{m}$ and the actual denoising can be directly applied as shown in the toy example in Fig \ref{fig:context-def}. Before we formally describe our DUDE-Seq for substitution-error correction, however, \textcolor{\proofcolor}{we need to address} some subtle points.

First, the original DUDE in (\ref{eq:dude_rule}) assumes that the DMC matrix $\mathbf{\Pi}$ is known beforehand, but in real DNA sequence denoising, we need to estimate $\mathbf{\Pi}$ for each sequencing device. As described in the Experimental Results section in detail, we \textcolor{\editagecolor}{performed this} estimation following the typical process \textcolor{\editagecolor}{for} obtaining the empirical confusion matrix\textcolor{\editagecolor}{,} \ie, \textcolor{\proofcolor}{we} align\textcolor{\editagecolor}{ed} the predefined reference sequence and its noise-corrupted sequence \textcolor{\proofcolor}{and} then \textcolor{\editagecolor}{determined} the ratio of substitution errors and obtain the estimated $\mathbf{\Pi}$.
Second, the original DUDE assumes that the noise mechanism is memoryless, \textcolor{\editagecolor}{\ie}, the error rate does not depend on the location of a base within the sequence.
In contrast, \textcolor{\editagecolor}{for} real sequencing devices, the actual error rate, namely, the conditional probability Pr$(Z_i=z|X_i=x)$ may not always be the same \textcolor{\editagecolor}{for all} location index \textcolor{\editagecolor}{values} $i$. For example, for Illumina sequencers, the error rate tends to increase towards the end\textcolor{\editagecolor}{s} of  reads\textcolor{\editagecolor}{,} as pointed out in \cite{laehnemann2015denoising}. In our DUDE-Seq, however, we still treat the substitution error mechanism as a DMC \textcolor{\proofcolor}{and therefore} use the single estimated $\mathbf{\Pi}$ obtained as above, which \textcolor{\proofcolor}{is} essentially \textcolor{\proofcolor}{the same as} \textcolor{\editagecolor}{that obtained} using the \emph{average} error rate matrix. Our experimental results show that such an approach still yields very competitive denoising results. Thirdly, the optimality of the original DUDE relies on the stationarity of the underlying clean sequence, \textcolor{\proofcolor}{thus} requir\textcolor{\proofcolor}{ing} a very large observation sequence length $n$ to obtain a reliable statistics vector $\mathbf{m}$. In contrast, most sequencing devices generate multiple short reads of lengths $100\sim200$. Hence, in DUDE-Seq, we combined all statistics vectors collected from multiple short reads to generate a single statistics vector $\mathbf{m}$ to use in (\ref{eq:dude_rule}).

\textcolor{\editagecolor}{Addressing} \textcolor{\proofcolor}{the} above three points, a formal summary of \acl~for the substitution errors is given in Algorithm~1. \textcolor{black}{Note that the pseudocode in Algorithm~1 skips those bases whose Phred quality score\textcolor{\proofcolor}{s are} higher than a user-specified threshold and invokes \acl~only for the bases with low quality scores (lines 10--14). This is in accord with the common practice in sequence preprocessing and \textcolor{\editagecolor}{is not a specific property of} the \acl~ algorithm.} Furthermore, for simplicity, we denoted $z^n$ as the entire noisy DNA sequence, and $\mathbf{m}^T(z^n,z_{i-k}^{i-1},z_{i+1}^{i+k})$ \textcolor{\editagecolor}{represents} the aggregated statistics vector obtained as described above.

\begin{algorithm*}[!ht]
    \caption{The \emph{DUDE-Seq} for substitution errors}\label{dude_alg}
    \begin{algorithmic}[1]
    	\small
    \Require Observation $z^n$, Estimated DMC matrix $\mathbf{\Pi}\in\mathbb{R}^{4\times4}$, Hamming loss $\mathbf{\Lambda}\in\mathbb{R}^{4\times4}$, Context size $k$, Phred quality score $Q^n$
    \Ensure The denoised sequence $\hat{X}^n$
    \State Define $\mathbf{m} (z^{n}, l^{k}, r^{k})\in\mathbb{R}^{4}$ for all $(l^k,r^k)\in\{\texttt{A}, \texttt{C}, \texttt{G}, \texttt{T}\}^{2k}$.
    \State Initialize $\mathbf{m} (z^{n}, l^{k}, r^{k})[a]=0$ for all $(l^k,r^k)\in\{\texttt{A}, \texttt{C}, \texttt{G}, \texttt{T}\}^{2k}$ and for all $a\in\{\texttt{A}, \texttt{C}, \texttt{G}, \texttt{T}\}$
    \For {$i \leftarrow k+1,\ldots, n-k$}
    \Comment \textsf{First pass}
        \State $\mathbf{m} (z^{n}, z_{i-k}^{i-1}, z_{i+1}^{i+k})[z_{i}] = \mathbf{m} (z^{n}, z_{i-k}^{i-1}, z_{i+1}^{i+k})[z_{i}] +1$
    \Comment \textsf{Update the count statistics vector}
    \EndFor
    \For {$i \leftarrow 1,\ldots, n$}
    \Comment \textsf{Second pass}
        \If {$i\leq k$ \texttt{or} $i\geq n-k+1$}
            \State $\hat{X}_i=z_i$
        \Else
            \If {$Q_i > \text{threshold}$}
            \Comment \textsf{Quality score}
            \State $\hat{X}_i=z_i$
            \Else
            \State $\hat{X}_i(z^n) =\argmin\limits_{\hat{x}\in\{\texttt{A},\texttt{C},\texttt{G},\texttt{T}\}}\mathbf{m}^T(z^n,z_{i-k}^{i-1},z_{i+1}^{i+k})\mathbf{\Pi}^{-1}[\lambda_{\hat{x}}\odot \pi_{z_i}]$
            \Comment \textsf{Apply the denoising rule}
            \EndIf
        \EndIf
    \EndFor
    \end{algorithmic}
\end{algorithm*}

\subsubsection*{Remarks.}
\begin{enumerate}
\item

\textcolor{\tsmploscolor}{Incorporating \textcolor{\editageploscolor}{flanking sequences} in DUDE-Seq is quite straightforward; we can simply use the one-sided contexts $l^{2k}$ or $r^{2k}$ once DUDE-Seq reaches the flank\textcolor{\editageploscolor}{ing} regions. In our experiments, however, we did not \textcolor{\editageploscolor}{perform} such modification (lines 7--8 of Algorithm~1) since we normally used small $k$ values (around $k=5$). As demonstrated in our experimental results, the effect of such small flank\textcolor{\editageploscolor}{ing regions} is not significant on the final denoising result\textcolor{\editageploscolor}{s}, and we can achieve satisfactory results without considering flank\textcolor{\editageploscolor}{ing regions}. However, in general, should longer values of $k$ be needed, we can easily modify the algorithm to incorporate one-sided contexts in the flank\textcolor{\editageploscolor}{ing} regions, and such modification will clearly improve the final denoising result.}

\item
\textcolor{\bhlploscolor}{\acl~does not need to consider reverse complements of the input sequences \textcolor{\editageploscolor}{to} collect $\mathbf{m}$'s, since forward and reverse reads \textcolor{\editageploscolor}{are handled} separately in our experiments. Reverse complements are typically considered when we need to handle double-stranded sequences without knowing whether each read corresponds to \textcolor{\editageploscolor}{the} forward or reverse strand.}
\end{enumerate}

\subsection*{Homopolymer Errors.}
\begin{figure}[!h]
	\centering
	\includegraphics[width=0.5\linewidth]{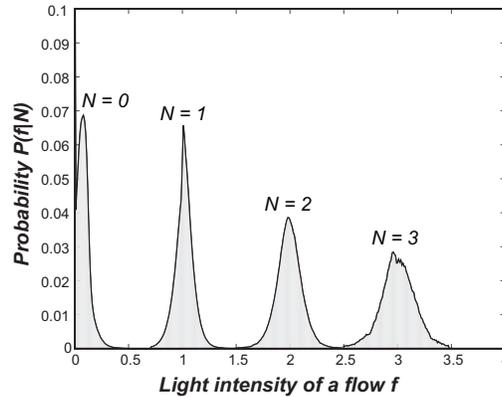}
	\caption{{\bf Conditional intensity distributions for $N = 0,1,2,3$.}}
	\label{fig:cont-disc-ch}
\end{figure}

\textcolor{\proofcolor}{H}omopolymer errors, \textcolor{\tsmrevisioncolor}{particularly in pyrosequencing,} occur while handling the observed flowgram, and a careful understanding of the error injection procedure is necessary to correct \textcolor{\editagecolor}{these errors}. As described in \cite{quince2011removing}, in pyrosequencing, the light intensities, \ie, flowgram, that correspond to a fixed order of four DNA bases $\{\texttt{T}, \texttt{A}, \texttt{C}, \texttt{G}\}$ are sequentially observed. The intensity value increases when the number of consecutive nucleotides (\ie, homopolymers) for each DNA base increases, and the standard base-calling procedure round\textcolor{\editagecolor}{s} the continuous-valued intensities to the closest integers. For example, when the observed light intensities for the two frames of DNA bases are $[0.03\; 1.03\; 0.09\; 0.12;\, 1.89\; 0.09\; 0.09\; 1.01],$ the corresponding rounded integers are $[0.00\; 1.00\; 0.00\; 0.00;\,2.00\;0.00\;0.00\;1.00]$. \textcolor{\editagecolor}{H}ence, the \textcolor{\editagecolor}{resulting} sequence is \texttt{ATTG}. The insertion and deletion errors \textcolor{\editagecolor}{are inferred} because the observed light intensities do not perfectly match the actual homopolymer lengths\textcolor{\proofcolor}{;} thus, the rounding procedure may \textcolor{\editagecolor}{result in the insertion or deletion of} DNA symbols. In fact, the distribution of the intensities $f$\textcolor{\proofcolor}{,} given the actual homopolymer length $N$, $\{P(f|N)\}$, can be obtained for each sequencing device, and Fig~\ref{fig:cont-disc-ch} shows typical distributions given various lengths.

Exploiting the fact that the order of DNA bases is always fixed \textcolor{\editagecolor}{at} $\{\texttt{T}, \texttt{A}, \texttt{C}, \texttt{G}\}$, we can apply \textcolor{black}{the setting of} the generalized DUDE in \cite{dembo2005universal} to correct homopolymer errors as follows. \textcolor{\proofcolor}{Because} we \textcolor{\proofcolor}{know} \textcolor{\editagecolor}{the} exact DNA base \textcolor{\editagecolor}{that corresponds with} each intensity value, the goal \textcolor{\editagecolor}{is the} correct estimat\textcolor{\editagecolor}{imation of} homopolymer lengths from the observed intensity values. Hence, we can interpret the intensity distributions $\{P(f|N)\}$ as the memoryless noisy channel models with \textcolor{\editagecolor}{a} continuous-output\textcolor{\editagecolor}{,} where the channel input is the homopolymer length $N$. We set the upper bound of $N$ to 9 \textcolor{\bhlploscolor}{according to the convention commonly used for handling flowgram distributions in the targeted amplicon sequencing literature~\cite{quince2011removing,bragg2013shining,fichot2013microbial}.}
When the usual rounding function
\be
Q_R(f) = \argmin_{i\in\{0,\ldots,9\}}|i-f|\label{eq:rounding}
\ee is used as a scalar quantizer\textcolor{\proofcolor}{,} as mentioned above, \textcolor{\proofcolor}{and} the virtual DMC $\mathbf{\Gamma}\in\mathbb{R}^{10\times 10}$ can be obtained \textcolor{\proofcolor}{by} calculating the integral
\be
\Gamma(i,j) = \int_{j-0.5}^{j+0.5}P(f|i)df\label{eq:gamma}
\ee
for each $0\leq i \leq 9, \ 1\leq j \leq 9$ and $\Gamma(i,0) = \int_{0}^{0.5}P(f|i)df$.

With this virtual DMC model, we apply a scheme inspired by the generalized DUDE to correctly estimate the homopolymer lengths, which results in correcting the insertion and deletion errors.
That is, we set $\mcX=\mcZ=\{0,1,\ldots,9\}$, and again use the Hamming loss $\mathbf{\Lambda}\in\mathbb{R}^{10\times 10}$. With this setting, we apply $Q_R(f)$ to each $f_i$ to obtain the quantized discrete output $z_i$, \textcolor{\editagecolor}{and} obtain the count statistics vector $\mathbf{m}$ from $z^n$ during the first pass. \textcolor{black}{Then, for the second pass, instead of applying the more involved denoising rule in \cite{dembo2005universal}, we employ the same rule as (\ref{eq:dude_rule}) with $\mathbf{\Gamma}$ in place of $\mathbf{\Pi}$ to obtain the denoised sequence of integers $\hat{X}^n$ based on the quantized noisy sequence $Z^n$. \textcolor{\editagecolor}{Although it is} potentially suboptimal compared to the generalized DUDE, this scheme \textcolor{\proofcolor}{is} used \textcolor{\tsmrevisioncolor}{\textcolor{\proofcolor}{because} \textcolor{\editagecolor}{its implementation is easier} and \textcolor{\editagecolor}{it} has \textcolor{\proofcolor}{a} faster running time than that of the generalized DUDE.}
Once we obtain $\hat{X}^n$, from the knowledge of the DNA base for each $i$, }we can reconstruct the homopolymer error-corrected DNA sequence $\hat{D}$ (the length of which may not necessarily be equal to $n$). Algorithm~2 summarizes the pseudo-code of \acl~for homopolymer-error \textcolor{\editagecolor}{correction}.

\begin{algorithm*}[!ht]
    \small
    \caption{The \emph{DUDE-Seq} for homopolymer errors}\label{dude_alg_homopolymer}
    \begin{algorithmic}[1]
    \Require Flowgram data $f^n$, Flowgram densities $\{P(f|N)\}_{N=0}^9$, Hamming loss $\mathbf{\Lambda}\in\mathbb{R}^{10\times10}$, Context size $k$
    \Ensure The denoised sequence $\hat{D}$
    \State Let $Q_R(f)$ be the rounding quantizer in Eq. (4) of the main text
    \State Let $\texttt{Base}(i)\in\{\texttt{T},\texttt{A},\texttt{C},\texttt{G}\}$ be the DNA base corresponding to $f_i$
    \State Define $\mathbf{m} (f^{n}, l^{k}, r^{k})\in\mathbb{R}^{10}$ for all $(l^k,r^k)\in\{0,1,\ldots,9\}^{2k}$.
    \State Initialize $\mathbf{m} (f^{n}, l^{k}, r^{k})[a]=0$ for all $(l^k,r^k)\in\{0,1,\ldots,9\}^{2k}$ and for all $a\in\{0,1,\ldots,9\}$
    \State Let $\hat{D}=\phi$, $I=0$
    \For {$i\leftarrow0,\ldots,9$}
    \For {$j\leftarrow0,\ldots,9$}
    \State Compute $\Gamma(i,j)$ following Eq. (5) of the main text
    \Comment \textsf{Computing the virtual DMC $\mathbf{\Gamma}$}
    \EndFor
    \EndFor
    \For {$i\leftarrow1,\ldots,n$} Obtain $z_i = Q_R(f_i)$
    \Comment \textsf{Note $z_i\in\{0,\ldots,9\}$}
    \EndFor
    \For {$i \leftarrow k+1,\ldots, n-k$}
    \Comment \textsf{First pass}
    \State $\mathbf{m} (f^{n}, z_{i-k}^{i-1}, z_{i+1}^{i+k})[z_{i}] = \mathbf{m} (f^{n}, z_{i-k}^{i-1}, z_{i+1}^{i+k})[z_{i}] +1$
    \EndFor
    \For {$i \leftarrow 1,\ldots, n$}
    \Comment \textsf{Second pass}
    \If {$i\leq k$ \texttt{or} $i\geq n-k+1$} $\hat{X}_i(f^n)=z_i$
    \Else
    \State $\hat{X}_i(f^n) =\argmin\limits_{\hat{x}\in\mcX}\mathbf{m}^T(f^n,z_{i-k}^{i-1},z_{i+1}^{i+k})\mathbf{\Gamma}^{-1}[\lambda_{\hat{x}}\odot \gamma_{z_i}]$
    \Comment \textsf{Note $\hat{X}_i(z^n)\in\{0,\ldots,9\}$}
    \EndIf
    \If {$\hat{X}_i(f^n)\geq1$}
    \For {$j\leftarrow 1,\ldots,\hat{X}_i(f^n)$} $\hat{D}_{I+j}= \texttt{Base}(i)$ \Comment \textsf{Reconstructing the DNA sequence}
    \EndFor
    \EndIf
    \State $I\leftarrow I+ \hat{X}_i(f^n)$
    \EndFor
    \end{algorithmic}
\end{algorithm*}

\section*{Experimental Results}\label{sec:experiments}
\subsection*{Setup.}
We used both real and simulated NGS datasets and compared the performance of \acl~with \textcolor{\editagecolor}{that of} several state-of-the-art error correction methods. \textcolor{\srycolor}{The list of alternative tools used for comparison and the rationale behind our choice\textcolor{\editagecolor}{s are} described in the next subsection.} \textcolor{\editagecolor}{When} the flowgram intensities of base-calling were available, we corrected both homopolymer and substitution errors\textcolor{\editagecolor}{;} otherwise\textcolor{\editagecolor}{, we} \textcolor{\proofcolor}{only} \textcolor{\editagecolor}{corrected} substitution errors. The specification\textcolor{\editagecolor}{s} of the machine we used \textcolor{\editagecolor}{for the analysis are} as follows: Ubuntu 12.04.3 LTS, 2$\times$ Intel Xeon X5650 CPUs, 64 GB main memory, and 2 TB HDD.

\begin{figure}[!h]
	\begin{adjustwidth}{-2in}{0in}
    	\centering
    	\includegraphics[width=\linewidth]{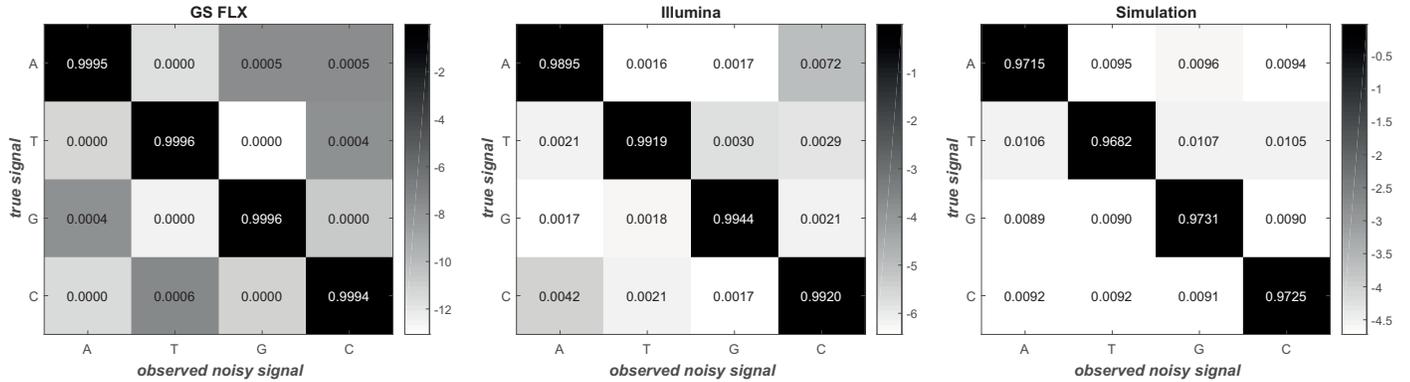}
    	\begin{spacing}{\mylinespacing}
    		\caption{\textcolor{\bhlplosrevisioncolor}{{\bf Adjustable DMC matrix $\mathbf{\Pi}$ of \acl.} Empirically obtained $\mathbf{\Pi}$'s for different sequencing platforms (colors are on a log scale).}}
    		\label{fig:results-pi}
    	\end{spacing}
	\end{adjustwidth}
\end{figure}

\acl~has a single hyperparameter $k$, the context size, that needs to be determined. \textcolor{\proofcolor}{S}imilar to the popular $k$-mer\textcolor{\editagecolor}{-}based schemes, there is no analytic\textcolor{\editagecolor}{al method for selecting} the best $k$ for finite data size $n$\textcolor{\proofcolor}{,} except for the asymptotic order result of $k|\mcX|^{2k}=o(n/\log n)$ in \cite{weissman2005universal}, but a heuristic rule of thumb is to try values between 2 and 8.
Furthermore, as shown in Eq. (\ref{eq:dude_rule}), the two adjustable matrices, \textcolor{\bhlcolor}{$\boldsymbol{\Lambda}$ and $\boldsymbol{\Pi}$,} are required for \acl. \textcolor{\bhlcolor}{The loss $\boldsymbol{\Lambda}$ used for both types of errors is the Hamming loss.}
\textcolor{\bhlploscolor}{According to Marinier~\etal~\cite{marinier2015pollux}, adjusting the sequence length by one can correct most homopolymer errors, which justifies our use of Hamming loss in \acl. In our experiments, \textcolor{\editageploscolor}{the use of} other types of loss functions did not \textcolor{\editageploscolor}{result in} any noticeable performance difference\textcolor{\editageploscolor}{s}.}
The DMC matrix $\boldsymbol{\Pi}$ for substitution errors is empirically determined by aligning each sampled read to its reference sequence\textcolor{\editagecolor}{,} as in \cite{quince2011removing}. \textcolor{\bhlplosrevisioncolor}{Fig~\ref{fig:results-pi}} shows the non-negligible variation \textcolor{\editagecolor}{in} the empirically obtained $\mathbf{\Pi}$'s across the sequencing platforms\textcolor{\editagecolor}{,} \textcolor{\bhlcolor}{where each row corresponds to the true signal $x$ and each column corresponds to the observed noisy signal $z$. In this setting, each cell represents the conditional probability $P(z|x)$. In our experiments, dataset P1--P8 used $\mathbf{\Pi}$ for GS FLX, Q19--Q31 used $\mathbf{\Pi}$ for Illumina, and S5, A5 used $\mathbf{\Pi}$ for Simulation \textcolor{\editagecolor}{data}. The details of each dataset \textcolor{\editagecolor}{are} explained in the following sections.}

In order to evaluate the results, we used Burrows-Wheeler Aligner (BWA)~\cite{li2009fast} and SAMtools~\cite{li2009sequence}. We aligned all reads to their reference genome using BWA with the following parameters: [minimum seed length: 19, matching score: 1, mismatch penalty: 4, gap open penalty: 6, gap extension penalty: 1].
\textcolor{\bhlploscolor}{After the mapped regions were determined \textcolor{\editageploscolor}{using} BWA in SAM format, we chose uniquely mapped pairs using SAMtools. The Compact Idiosyncratic Gapped Alignment Report (CIGAR) string and MD
tag (string for mismatching positions) \textcolor{\editageploscolor}{for} each of the resultant pairs in the SAM file were reconstructed to their pairwise alignments using sam2pairwise~\cite{lafave_2014_11377}.}


\subsection*{\textcolor{\bhlplosrevisioncolor}{Evaluation Metric.}}
\textcolor{\bhlplosrevisioncolor}{As a performance measure, we define the per-base error rate of a tool after denoising as
\begin{align}
e_\text{tool} = \frac{\text{\# mismatched bases}}{\text{\# aligned bases}},\label{eq:error_rate_def}
\end{align}
in which `\# aligned bases' represents the number of mapped bases (\ie, matches and mismatches) after mapping each read to its reference sequence, and `\# mismatched bases' represents the number of the erroneous bases (\ie, insertions, deletions, and substitutions) among the aligned bases.}

\textcolor{\bhlplosrevisioncolor}{We also employ an alternative definition that adjusts the error rate by incorporating the degree of alignment. To this end, we define the \emph{relative gain} of the number of aligned bases after denoising by a tool over raw data as
\begin{align}
    g(a_\text{tool}) = \frac{\text{\# aligned bases after denoising}-\text{\# aligned bases in raw}}{\text{\# aligned bases in raw}}. \label{eq:rel_gain_align}
\end{align}
Based on this, the adjusted error rate $\hat{e}_\text{tool}$ of a denoising tool is defined as follows:
\begin{align}
    \hat{e}_{\text{tool}} = (1+g(a_\text{tool})) \times e_{\text{tool}} - g(a_\text{tool}) \times e_{\text{raw}},\label{eq:error_hat}
\end{align}
where $e_\text{tool}$ and $e_\text{raw}$ represent the (unadjusted) error rates of the denoised data and the raw data, respectively. In other words, (\ref{eq:error_hat}) is a weighted average of $e_\text{tool}$ and $e_\text{raw}$, in which the weights are determined by the relative number of aligned bases of a tool compared to the raw sequence. We believe $\hat{e}_{\text{tool}}$ is a fairer measure as it penalizes the error rate of a denoiser when there is a small number of aligned bases. The relative gain of the adjusted error rate over raw data is then defined as
\begin{align}
    g(\hat{e}_{\text{tool}}) = \frac{e_{\text{raw}} - \hat{e}_{\text{tool}}}{e_{\text{raw}}},\label{eq:rel_gain_error}
\end{align}
which we use to evaluate the denoiser performance.}

\textcolor{\bhlplosrevisioncolor}{While evaluating a clustering result, we employ a measure of concordance (MoC)~\cite{pfitzner2009characterization} which is a popular similarity measure for pairs of clusterings.
For two pairs of clusterings $P$ and $Q$ with $I$ and $J$ clusters, respectively, the MoC is defined as
\begin{align}
    \text{MoC} (P,Q) = \frac{1}{\sqrt{IJ}-1} \left (\sum_{i=1}^{I} \sum_{j=1}^{J} \frac{f_{ij}^{2}}{p_{i} q_{j}}-1 \right )
\end{align}
where $f_{ij}$ is the number of the common objects between cluster $P_i$ and $Q_j$ when $p_i$ and $q_j$ are the numbers of the objects in cluster $P_i$ and $Q_j$, respectively.
A MoC of one or zero represents perfect or no concordance, respectively, between the two clusters.}

\subsection*{\textcolor{\bhlploscolor}{Software Chosen for Comparison.}}
It \textcolor{\editagecolor}{is} impossible to compare the performance of DUDE-Seq with \textcolor{\editagecolor}{that of} all other schemes. Hence, we selected representative baselines \textcolor{\proofcolor}{using} the following reasoning\textcolor{\editagecolor}{.}
\begin{enumerate}
	\item We included tools that can represent different principles outlined in the Introduction\textcolor{\editagecolor}{,} namely, $k$-mer\textcolor{\editagecolor}{-}based (Trowel, Reptile, BLESS, and fermi), MSA\textcolor{\proofcolor}{-}based (Coral), and statistical error model\textcolor{\proofcolor}{-}based (AmpliconNoise) \textcolor{\editagecolor}{methods}.
	\item We cons\textcolor{\proofcolor}{idered} the recommendation\textcolor{\proofcolor}{s} of~\cite[Table 2]{laehnemann2015denoising} to choose baseline tools that are competitive for different scenarios\textcolor{\proofcolor}{,} \textcolor{\editagecolor}{\ie,} for 454 pyrosequencing data (AmpliconNoise), non-uniform coverage data\textcolor{\editagecolor}{,} such as metagenomics \textcolor{\editagecolor}{data} (Trowel, fermi, Reptile), data \textcolor{\editagecolor}{dominated by} substitution errors\textcolor{\editagecolor}{,} such as Illumina data (Trowel, fermi, Reptile), and data \textcolor{\proofcolor}{with} \textcolor{\editagecolor}{a high prevalence of} indel errors (Coral).
	\item For multiple $k$-mer\textcolor{\editagecolor}{-}based tools, we chose \textcolor{\editagecolor}{those} that use different main approaches/data structures: BLESS ($k$-mer spectrum\textcolor{\editagecolor}{-}based/hash table and bloom filter), fermi ($k$-mer spectrum and frequency\textcolor{\editagecolor}{-}based/hash table and suffix array), Trowel ($k$-mer spectrum\textcolor{\editagecolor}{-}based/hash table), and Reptile ($k$-mer frequency and Hamming graph\textcolor{\editagecolor}{-}based/replicated sorted $k$-mer list).
	\item The selected tools were developed quite recently; Trowel and BLESS (2014), fermi (2012), Coral and AmpliconNoise (2011), and Reptile (2010).
	\item We mainly chose tools that return read-by-read denoising results to make fair error-rate comparisons \textcolor{\sryoonrevisioncolor}{with DUDE-seq}. \textcolor{\editagecolor}{W}e excluded tools that return a substantially reduced number of reads after error correction (caused by filtering or forming consensus clusters). \textcolor{\proofcolor}{E}xamples of excluded tools are Acacia, ALLPATHS-LG, and SOAPdenovo.
	\item We also excluded some recently developed tools that require additional mandatory information (e.g., the size of the genome of the reference organism) beyond the common setting of DNA sequence denoising \textcolor{\proofcolor}{in order} to make fair error-rate comparisons. \textcolor{\proofcolor}{E}xamples of excluded tools are Fiona, Blue, and Lighter. Incorporating those \textcolor{\editagecolor}{tools that require} additional information \textcolor{\proofcolor}{in}to \textcolor{\proofcolor}{the} DUDE-Seq framework and \textcolor{\editagecolor}{comparisons} with the excluded tools would be another future direction\textcolor{\editagecolor}{s}.
\end{enumerate}

\subsection*{Real Data: 454 Pyrosequencing.}
Pyrosequenced 16S rRNA genes are commonly used to characterize microbial communities \textcolor{\editagecolor}{because the method yields} relatively longer reads than those of other NGS technologies~\cite{reeder2010rapid}. \textcolor{\sryoonrevisioncolor}{Although 454 pyrosequencing is gradually being phased out, we test\textcolor{\editagecolor}{ed} \acl~with 454 pyrosequencing data \textcolor{\editagecolor}{for} the following \textcolor{\editagecolor}{reasons}: (1) the \acl~methodology for correcting homopolymeric errors in 454 sequencing \textcolor{\editagecolor}{data} is equally applicable to other sequencing technologies that produce homopolymeric errors, such as Ion Torrent; (2) using pyrosequencing data allows us to exploit existing (experimentally obtained) estimates of the channel transition matrix $\mathbf{\Gamma}$ (\eg, \cite{quince2011removing}), which is required for denoising noisy flowgrams by \acl~\textcolor{\bhlcolor}{(see Algorithm 2)}; (3) in the metagenomics literature, widely used standard benchmarks consist of datasets generated by pyrosequencing.}

In metagenome analysis~\cite{schloss2005introducing}, grouping reads and assigning them to operational taxonomic units (OTUs) (\ie, binning) are essential processes\textcolor{\editagecolor}{,} \textcolor{\bhlploscolor}{given that the majority of microbial species have not been taxonomically classified.}
\textcolor{\bhlploscolor}{By OTU binning, we can computationally identify closely related genetic groups of reads at a desired level of sequence differences.}
However, \textcolor{\editagecolor}{owing} to erroneous reads, \textcolor{\proofcolor}{non}existent OTUs may \textcolor{\editagecolor}{be obtained}, \textcolor{\editagecolor}{resulting} in the common problem of overestimating ground truth OTUs. Such overestimation \textcolor{\editagecolor}{is a} bottleneck \textcolor{\editagecolor}{in} the overall microbiome analysis\textcolor{\editagecolor}{;} hence, removing errors \textcolor{\editagecolor}{in} reads before \textcolor{\editagecolor}{they are assigned} to OTUs \textcolor{\editagecolor}{is} a critical issue \cite{quince2011removing}. With this motivation, in some of our experiments below, we used the difference \textcolor{\editagecolor}{between} the number of assigned OTUs and the ground truth number of OTUs as a proxy \textcolor{\editagecolor}{for} denoising performance; \textcolor{\bhlploscolor}{the number of OTUs \textcolor{\sryploscolor}{was} determined \textcolor{\editageploscolor}{using} UCLUST~\cite{edgar2010search}} \textcolor{\bhlplosrevisioncolor}{at identity threshold of 0.97 which is for species assignment.}

We tested the performance of \acl~with the eight datasets used in~\cite{quince2011removing}, which are mixtures of \textcolor{\bhlplosrevisioncolor}{94 environmental clones library from eutrophic lake (Priest Pot) using primers 787f and 1492r}. Dataset P1 ha\textcolor{\editagecolor}{d} 90 clones that are mixed in two orders of magnitude difference while P2 ha\textcolor{\editagecolor}{d} 23 clones that \textcolor{\editagecolor}{were} mixed in equal proportions. In P3, P4, \textcolor{\proofcolor}{and} P5 and P6, P7, \textcolor{\proofcolor}{and} P8, there are 87 mock communities mixed in even and uneven proportions, respectively. In all datasets, both homopolymer and substitution errors exist, and the flowgram intensity values as well as the distributions are available~\cite{quince2011removing}. Therefore, \acl~tries to correct both types of errors using the empirically obtained $\mathbf{\Pi}$ and the flowgram intensity distributions $\{P(f|N)\}$.

\begin{figure}[!h]
	\begin{adjustwidth}{-2in}{0in}
    	\centering
    	\includegraphics[width=\linewidth]{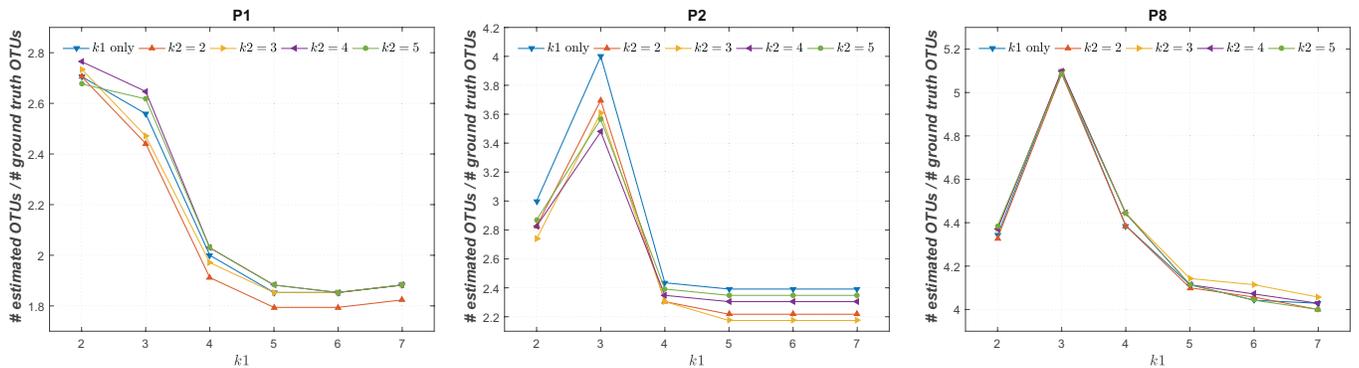}
    	\begin{spacing}{\mylinespacing}
    		\caption{\textcolor{\bhlplosrevisioncolor}{{\bf Hyperparameter $k$ of \acl.} Effects of varying context size $k$ [$k1$ is for Algorithm 1 (substitution-error correction) and $k2$ is for Algorithm 2 (homopolymer-error correction); data:~\cite{quince2011removing}].}}
    		\label{fig:results-k}
    	\end{spacing}
	\end{adjustwidth}
\end{figure}

We first show the effect of $k$ on the performance of \acl~in \textcolor{\bhlplosrevisioncolor}{Fig~\ref{fig:results-k}}. The vertical axis shows the \textcolor{\bhlplosrevisioncolor}{ratio} between the number of OTUs assign\textcolor{\editagecolor}{ed} after denoising with \acl~and the ground truth number of OTUs for the \textcolor{\bhlcolor}{P1, P2, and P8} dataset. The horizontal axis \textcolor{\proofcolor}{shows} the $k$ values used for correcting the substitution errors \textcolor{\bhlcolor}{(\ie, for Algorithm~1)}, and color-coded curves \textcolor{\editagecolor}{were generated} for different $k$ values used for homopolymer-error \textcolor{\editagecolor}{correction} \textcolor{\bhlcolor}{(\ie, for Algorithm~2)}. \textcolor{\editagecolor}{As shown in the figure,} correcting homopolymer errors (\ie, with $k=2$ for Algorithm~2) always enhance\textcolor{\editagecolor}{ed} the results in terms of the number of OTUs \textcolor{\proofcolor}{in comparison to} correcting substitution errors alone (\ie, Algorithm~1 alone). \textcolor{\editagecolor}{W}e observe that $k=5$ for Algorithm~1 and $k=2$ for Algorithm~2 \textcolor{\proofcolor}{produce} the best result\textcolor{\editagecolor}{s} in terms of the number of OTUs. \textcolor{\editagecolor}{L}arger $k$ \textcolor{\proofcolor}{value} work better for substitution errors \textcolor{\editagecolor}{owing to} the smaller alphabet size of the data, \ie, 4, compared to that of homopolymer errors, \ie, 10. \textcolor{black}{Motivated by this result, we fixed the context sizes of substitution error correction and homopolymer error correction to $k=5$ and $k=2$, respectively, for all subsequent experiments.}

\begin{figure}[!h]
	\begin{adjustwidth}{-2in}{0in}
		\centering
		\includegraphics[width=\linewidth]{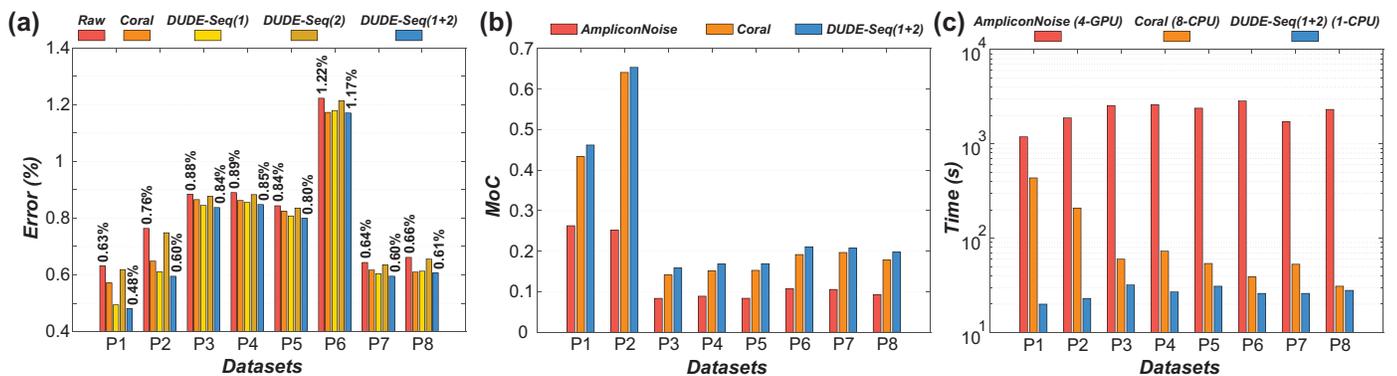}
		\begin{spacing}{\mylinespacing}
			\caption{{\bf Comparison of reads correction performance on eight real 454 pyrosequencing datasets (labeled P1--P8; \cite{quince2011removing}).} {[parameters: $k=5$ \textcolor{\sryoonrevisioncolor}{(Algorithm 1) and $k=2$ (Algorithm 2)} for \acl; $(s_{\text{PyroNoise}},c_{\text{PyroNoise}},s_{\text{SeqNoise}},c_{\text{SeqNoise}})=(60,0.01,25,0.08)$ for AmpliconNoise; $(k,mr,mm,g)=(21,2,2,3)$ for Coral]}: (a) Per-base error rates [1 and 2 represents substitution error-correction \textcolor{\bhlcolor}{(Algorithm~1)} and homopolymer error-correction \textcolor{\bhlcolor}{(Algorithm~2)}, respectively.]
				(b) Measure of concordance (MoC), a similarity measure for pairs of clusterings (c) Running time {(the type and quantity of processors used for each case are shown in legend)}}
			\label{fig:results-pyrosequencing}
		\end{spacing}
	\end{adjustwidth}
\end{figure}

In Fig~\ref{fig:results-pyrosequencing}(a), we report a more direct \textcolor{\editagecolor}{analysis of} error correction performance.
We compare\textcolor{\proofcolor}{d} the performance of \acl~with \textcolor{\editagecolor}{that of} Coral~\cite{salmela2011correcting}, which is \textcolor{\editagecolor}{an} \textcolor{\tsmcolor}{MSA-based} state-of-the-art scheme. It aligns multiple reads by exploiting \textcolor{\proofcolor}{the} $k$-mer neighborhood of each base read and produces read-by-read correction results for pyrosequencing datasets\textcolor{\editagecolor}{, similar to} \acl. \textcolor{\tsmrevisioncolor}{Furthermore, as a baseline, we also  present\textcolor{\proofcolor}{ed} the error rates \textcolor{\editagecolor}{for} the original, uncorrected sequences (labeled `Raw').} {\textcolor{\editagecolor}{W}e did not include \textcolor{\editagecolor}{the results of} AmpliconNoise~\cite{quince2011removing}, \textcolor{\tsmrevisioncolor}{a state-of-the-art scheme for 454 pyrosequencing \textcolor{\editagecolor}{data},} in \textcolor{\tsmrevisioncolor}{the} performance comparison \textcolor{\proofcolor}{because} it does not provide read-by-read correction results\textcolor{\editagecolor}{,} making a fair comparison \textcolor{\tsmcolorfinal}{\textcolor{\editagecolor}{of} the per-base error correction performance} with \acl~difficult.}
\textcolor{\editagecolor}{W}e observe\textcolor{\editagecolor}{ed} that \acl(1+2), which corrects both substitution errors and homopolymer errors, always outperforms Coral, and the relative error reductions of \acl(1+2) \textcolor{\editagecolor}{with respect to} `Raw\textcolor{\editagecolor}{,}' without any denoising\textcolor{\editagecolor}{, was} up to 23.8\%. Furthermore, the homopolymer error correction further drive\textcolor{\editagecolor}{s} down the error rates \textcolor{\editagecolor}{obtained by} substitution-error correction alone\textcolor{\proofcolor}{; hence,} \acl(1+2) always \textcolor{\editagecolor}{outperforms} \acl(1).

\textcolor{\tsmcolor}{
In Fig~\ref{fig:results-pyrosequencing}(b), we compare the error correction performance of three schemes, AmpliconNoise, Coral, and \acl, in terms of the \textcolor{\bhlplosrevisioncolor}{MoC}.} \textcolor{\tsmrevisioncolor}{AmpliconNoise assumes a certain statistical model on the DNA sequence and runs an expectation-maximization algorithm for denoising.}
Here, the two clusterings \textcolor{\editagecolor}{in the comparison} are the golden OTU clusterings and the clusterings returned by denoisers.
\textcolor{\tsmcolor}{We observe that for all eight datasets, the number of OTUs generated by \acl~is consistently closer to the ground truth, \textcolor{\proofcolor}{providing} higher MoC values than \textcolor{\editagecolor}{those of} the other two schemes.}

\textcolor{\tsmcolor}{Furthermore, Fig~\ref{fig:results-pyrosequencing}(c) compares the running time of the three schemes \textcolor{\editagecolor}{for} the eight datasets. We can clearly see that \acl~is substantially faster than the other two. Particularly, we stress that the running time of \acl, even when {implemented and executed with} a single CPU, is two orders of magnitude faster than that of {parallelized AmpliconNoise, \textcolor{\editagecolor}{run} on four powerful GPUs}. 
We believe \textcolor{\editagecolor}{that this substantial} boost over state-of-the-art scheme\textcolor{\proofcolor}{s} \textcolor{\editagecolor}{with respect to} running time \textcolor{\editagecolor}{is} a compelling reason for \textcolor{\proofcolor}{the} adoption of \acl~in microbial community analysis.}

\subsection*{Real Data: Illumina Sequencing.}\label{sec:real_illumina}
\textcolor{\tsmcolor}{Illumina platforms, such as GAIIx, MiSeq, and HiSeq, are \textcolor{\editagecolor}{currently} ubiquitous platforms in genome analysis. These platforms intrinsically generate paired-end reads (forward and reverse reads), due to the relatively short reads compared to \textcolor{\editagecolor}{those obtained by} automated Sanger sequencing~\cite{bartram2011generation}. Merging the forward and reverse reads from paired-end sequencing \textcolor{\editagecolor}{yeilds} elongated reads (\eg, $2\times300$ bp for MiSeq) that improve the performance \textcolor{\editagecolor}{of} downstream pipelines~\cite{magovc2011flash}.}

\textcolor{\tsmcolor}{Illumina platforms primarily inject substitution errors. A realistic error model is not the DMC, though, as the error rates of the Illumina tend to increase from the beginning to the end of reads\textcolor{\editagecolor}{. T}hus, the assumptions under which the DUDE was originally developed do not exactly apply to the error model of Illumina. In our experiments with \acl, however, we still used the empirically obtained DMC model $\mathbf{\Pi}$ in \textcolor{\bhlplosrevisioncolor}{Fig~\ref{fig:results-pi}}, which was computed by \emph{averaging} all error rates throughout different Illumina platforms.}

\begin{figure}[!h]
	\begin{adjustwidth}{-2in}{0in}
		\centering
		\includegraphics[width=0.95\linewidth]{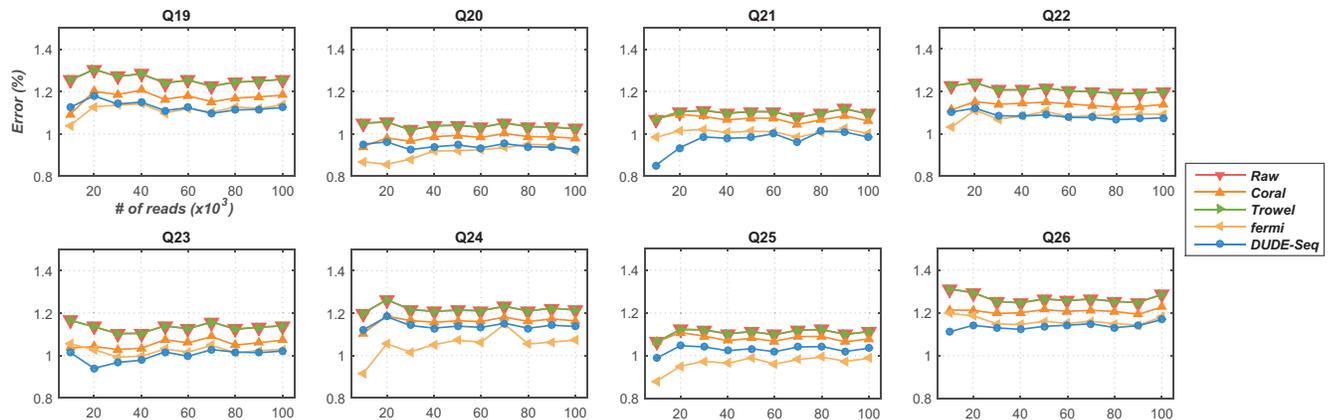}
		\begin{spacing}{\mylinespacing}
			\caption{{\bf Comparison of reads correction performance on real Illumina datasets (labeled Q19--Q26; \textcolor{\sryoonrevisioncolor}{see Table~\ref{tab:illumina-dataset} for more details}).} [parameters: $(k,mr,mm,g)=(21,1,1,1000)$ for Coral; $k=21$ for Trowel; $(k,O,C,s)=(21,3,0.3,5)$ for fermi; $k=5$ for \acl; \textcolor{\sryoonrevisioncolor}{no BLESS result shown since it did not work on these data] [Organisms: \textit{Anaerocellum thermophilum Z-1320 DSM 6725} (Q19 and Q23), \textit{Bacteroides thetaiotaomicron VPI-5482} (Q20 and Q24), \textit{Bacteroides vulgatus ATCC 8482} (Q21 and Q25), \textit{Caldicellulosiruptor saccharolyticus DSM 8903} (Q22 and Q26)] [Q19--Q22: Miseq (Library: nested single index, Taq: Q5 neb, Primer: 515 \& 805RA)] [Q23--Q26: Miseq (Library: NexteraXT, Taq: Q5 neb, Primer: 341f \& 806rcb)]}}
			\label{fig:results-illumina}
		\end{spacing}
	\end{adjustwidth}
\end{figure}

\textcolor{\sryoonrevisioncolor}{
In our experiments, we used 13} real Illumina datasets (named Q19--Q31) reported \textcolor{\editagecolor}{previously}~\cite{schirmer2015insight}, \textcolor{\sryoonrevisioncolor}{\textcolor{\editagecolor}{including} sequencing results from} four organisms (\textit{Anaerocellum thermophilum Z-1320 DSM 6725}, \textit{Bacteroides thetaiotaomicron VPI-5482}, \textit{Bacteroides vulgatus ATCC 8482}, and \textit{Caldicellulosiruptor saccharolyticus DSM 8903}) target\textcolor{\editagecolor}{ing} two hypervariable regions\textcolor{\editagecolor}{,} V3 and V4\textcolor{\editagecolor}{,} using different configurations (\textcolor{\tsmrevisioncolor}{\textcolor{\editagecolor}{s}ee the caption for Table~\ref{tab:illumina-dataset} and Fig~\ref{fig:results-illumina} for details). }
To \textcolor{\editagecolor}{examine} how the number of reads in a dataset affects denoising performance, we derived 10 subsets from the original datasets by randomly subsampling 10,000 to 100,000 reads in \textcolor{\editagecolor}{increments} of 10,000 reads. In addition to Coral, we compared the performance of \acl~with \textcolor{\editagecolor}{that of} BLESS~\cite{heo2014bless}, fermi~\cite{li2012exploring}, and Trowel~\cite{lim2014trowel}, \textcolor{\sryoonrevisioncolor}{which are \textcolor{\tsmrevisioncolor}{representative} $k$-mer\textcolor{\editagecolor}{-}based state-of-the-art tools.}
BLESS corrects ``weak'' $k$-mers that exist between consecutive ``solid'' $k$-mers, assuming that a weak $k$-mer has only one error. \textcolor{\tsmrevisioncolor}{F}ermi corrects sequencing errors in underrepresented $k$-mers using a heuristic cost function based on quality scores and does not rely on a $k$-mer occurrence threshold.
Trowel does not use a coverage threshold for its $k$-mer spectrum and iteratively boosts the quality values of bases after making corrections with $k$-mers that have high quality values.

\setlength{\tabcolsep}{13pt}
\ctable[
caption = {\textcolor{\srycolor}{\bf Details of the Illumina datasets~\cite{schirmer2015insight} used for our experiments shown in Fig ~\ref{fig:results-illumina}}},
label = {tab:illumina-dataset},
    doinside = \small,
    pos = !ht,
]
{cccccc}
{
	\tnote[]{\scriptsize{Taqs: HiFI Kapa (HF), Q5 neb (Q5); Organisms: Anaerocellum thermophilum Z-1320 DSM 6725 (AT), Bacteroides thetaiotaomicron VPI-5482 (BT), Bacteroides vulgatus ATCC 8482 (BV), Caldicellulosiruptor saccharolyticus DSM 8903 (CS), Herpetosiphon aurantiacus ATCC 23779 (HA), Rhodopirellula baltica SH 1 (RBS), Leptothrix cholodnii SP-6 (LC)}}
}
{
	\toprule
	dataset & \multirow{2}[0]{*}{region} & \multirow{2}[0]{*}{sequencer} & \multirow{2}[0]{*}{Taq} & \multirow{2}[0]{*}{organism} & \textcolor{\srycolor}{forward \& reverse} \\
	ID    &       &       &       &       & primer \\
	\midrule
	Q19    & V4    & MiSeq2 & Q5    & AT    & 515 \& 805RA \\
	Q20    & V4    & MiSeq2 & Q5    & BT    & 515 \& 805RA \\
	Q21    & V4    & MiSeq2 & Q5    & BV    & 515 \& 805RA \\
	Q22    & V4    & MiSeq2 & Q5    & CS    & 515 \& 805RA \\
	Q23    & V4    & MiSeq2 & HF    & AT    & 515 \& 805RA \\
	Q24    & V4    & MiSeq2 & HF    & BT    & 515 \& 805RA \\
	Q25    & V4    & MiSeq2 & HF    & BV    & 515 \& 805RA \\
	Q26    & V4    & MiSeq2 & HF    & CS    & 515 \& 805RA \\
	Q27    & V3/V4 & MiSeq1 & Q5    & AT    & 314f \& 806rcb \\
	Q28    & V3/V4 & MiSeq1 & Q5    & BT    & 314f \& 806rcb \\
	Q29    & V3/V4 & MiSeq1 & Q5    & BV    & 314f \& 806rcb \\
	Q30    & V3/V4 & MiSeq1 & Q5    & CS    & 314f \& 806rcb \\
	Q31    & V3/V4 & MiSeq1 & HF    & AT    & 314f \& 806rcb \\
	\bottomrule
}

Fig~\ref{fig:results-illumina} shows the per-base \textcolor{\tsmrevisioncolor}{error rates, defined in (\ref{eq:error_rate_def}),}
\textcolor{\sryoonrevisioncolor}{\textcolor{\editagecolor}{for} the tools under comparison
} \textcolor{\editagecolor}{using} the \textcolor{\tsmrevisioncolor}{first} eight datasets (Q19--Q26) \textcolor{\sryoonrevisioncolor}{and their subsets created as described above (\textcolor{\tsmrevisioncolor}{thus,} a total of 80 datasets per tool).}
\textcolor{\proofcolor}{BLESS did not run \textcolor{\editagecolor}{successfully} on these datasets, \textcolor{\editagecolor}{and} hence its result\textcolor{\editagecolor}{s are} not shown.}
\textcolor{\sryoonrevisioncolor}{\textcolor{\tsmrevisioncolor}{First, w}e can confirm that \acl~is effective in reducing substitution error\textcolor{\tsmrevisioncolor}{s for \textcolor{\editagecolor}{data obtained using} the Illumina} \textcolor{\tsmrevisioncolor}{platform} in all tested cases \textcolor{\tsmrevisioncolor}{of targeted amplicon sequencing, \textcolor{\proofcolor}{with} relative error rate reductions of} 6.40--49.92\%\textcolor{\proofcolor}{,} compared to the `Raw' sequences. \textcolor{\tsmrevisioncolor}{Furthermore, among the tools \textcolor{\editagecolor}{included in the comparison},} \acl~\textcolor{\proofcolor}{produced} the best results for the largest number of datasets. For Q24 and Q25, fermi was most effective\textcolor{\editagecolor}{,} but was outperformed by \acl~in many other cases. Coral was able to denoise to some extent but was \textcolor{\bhlploscolor}{inferior} to \acl~and fermi. Trowel gave unsatisfactory results in this experiment.}

\textcolor{\tsmrevisioncolor}{Before presenting our next result\textcolor{\editagecolor}{s}, we note that while the error rate defined in (\ref{eq:error_rate_def}) is widely used for DNA sequence denoising research as a performance measure, it \textcolor{\editagecolor}{occasionally} misleading and \textcolor{\editagecolor}{can}not \textcolor{\editagecolor}{be used to} fairly evaluate the performance of denoisers. \textcolor{\editagecolor}{This} is because only errors \textcolor{\editagecolor}{at} aligned bases are counted in the error rate calculation; hence, a poor denoiser may significantly reduce the number of aligned bases, potentially further corrupting the noisy sequence, \textcolor{\proofcolor}{but it} can have \textcolor{\proofcolor}{a} low error rate calculated as in (\ref{eq:error_rate_def}). In our experiments with the datasets Q27-Q31, we \textcolor{\editagecolor}{detected} \textcolor{\proofcolor}{a} large variance \textcolor{\editagecolor}{in} the number of aligned bases across different denoising tools\textcolor{\proofcolor}{;} thus, it was \textcolor{\proofcolor}{difficult} to make a fair comparison among the performance of different tools with (\ref{eq:error_rate_def}).
}
\textcolor{\proofcolor}{We note that \textcolor{\editagecolor}{in} the experiments presented in Fig~\ref{fig:results-pyrosequencing}(a) and Fig~\ref{fig:results-illumina}\textcolor{\editagecolor}{,} such \textcolor{\editagecolor}{a} large variance \textcolor{\editagecolor}{was not detected}.}
\textcolor{\bhlplosrevisioncolor}{To alleviate this issue, we employ the alternative definition of the per-base error rate of a tool in Eq.~(\ref{eq:error_hat}).}

\begin{figure}[!h]
	\begin{adjustwidth}{-2in}{0in}
		\centering
		\includegraphics[width=\linewidth]{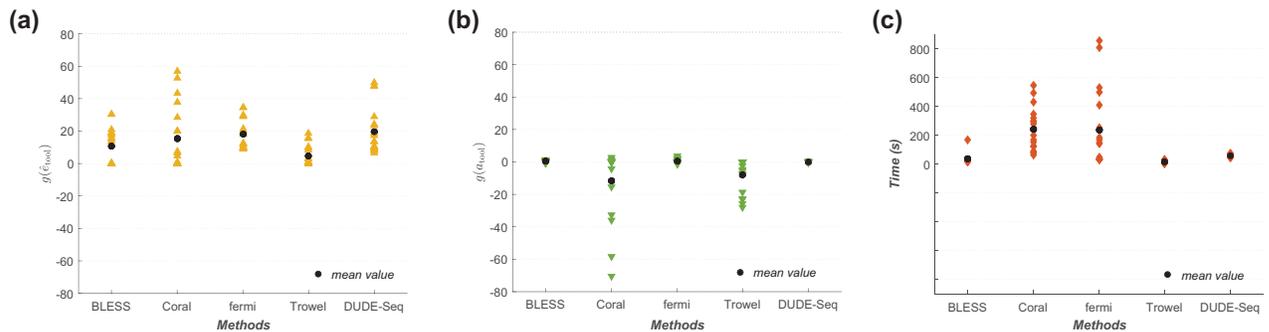}
		\begin{spacing}{\mylinespacing}
			\caption{{\bf Performance comparison.} (a) Relative gain of adjusted error rates over `Raw' data (Eq.~\ref{eq:rel_gain_error}). (b) Relative gain of aligned bases (Eq.~\ref{eq:rel_gain_align}). (c) Running time on real Illumina datasets (labeled Q19--Q31; \textcolor{\sryoonrevisioncolor}{see the caption for Fig~\ref{fig:results-illumina}).} [parameters: $\text{kmerlength}=21$ for BLESS; $(k,mr,mm,g)=(21,1,1,1000)$ for Coral; $k=21$ for Trowel; $(k,O,C,s)=(21,3,0.3,5)$ for fermi; $k=5$ for \acl] \textcolor{\sryoonrevisioncolor}{[BLESS did not work on Q19--Q26]}}
			\label{fig:results-weighted-gain}
		\end{spacing}
	\end{adjustwidth}
\end{figure}

\textcolor{\sryoonrevisioncolor}{
\textcolor{\tsmrevisioncolor}{Fig~\ref{fig:results-weighted-gain} shows the results \textcolor{\editagecolor}{obtained for} 100,000-read subsets of each of the Q19--Q31 datasets\textcolor{\tsmrevisioncolor}{, i.e., \textcolor{\editagecolor}{all} datasets,} for \acl~and the four alternative denoisers. \textcolor{\proofcolor}{Because} the datasets Q27--Q31 had two subsets of 100,000 reads, we used a total of 18 datasets to draw Fig~\ref{fig:results-weighted-gain}\textcolor{\proofcolor}{,} one each from Q19--Q26 \textcolor{\proofcolor}{and} two each from Q27--Q31. As mentioned \textcolor{\editagecolor}{previously}, BLESS could not \textcolor{\editagecolor}{run} successfully on Q19--Q26\textcolor{\proofcolor}{;} hence, there are only 10 points for BLESS in the plots.}
Fig~\ref{fig:results-weighted-gain}(a), (b) and (c) presents the distribution of $g(\hat{e}_{\text{tool}})$, $g(a_{\text{tool}})$, and running time\textcolor{\editagecolor}{s} for each tool, respectively. For each distribution, the average value is marked with a solid circle. \textcolor{\tsmrevisioncolor}{\textcolor{\editagecolor}{As shown in} Fig~\ref{fig:results-weighted-gain}(b), we clearly see that Coral and Trowel show \textcolor{\proofcolor}{a} large variance \textcolor{\proofcolor}{in} the number of aligned bases. For example, Coral only aligns 30\% of bases compared to \textcolor{\proofcolor}{the} raw sequence after denoising for some dataset\textcolor{\proofcolor}{s}. With the effect of this variance \textcolor{\proofcolor}{in} aligned bases adjusted, Fig~\ref{fig:results-weighted-gain}(a) shows that \acl~\textcolor{\proofcolor}{produces} the highest average $g(\hat{e}_{\text{tool}})$, \ie, 19.79\%, among all \textcolor{\proofcolor}{the} compared tools. Furthermore,} the variability of $g(a_{\text{tool}})$ was the smallest for \acl\textcolor{\editagecolor}{, as shown} in Fig~\ref{fig:results-weighted-gain}(b), suggesting its robustness.
\textcolor{\tsmrevisioncolor}{
Finally, \textcolor{\proofcolor}{in} Fig~\ref{fig:results-weighted-gain}(c), we observe that the running time\textcolor{\editagecolor}{s were significantly shorter for} \acl~and Trowel than \textcolor{\editagecolor}{for} Coral and fermi. Overall, we can conclude that DUDE-Seq is the most robust tool\textcolor{\editagecolor}{,} \textcolor{\proofcolor}{with a} fast running time and the highest average accuracy after denoising.}
}

\textcolor{\tsmrevisioncolor}{In summary,} we observe from Fig~\ref{fig:results-illumina} and Fig~\ref{fig:results-weighted-gain} that \acl~robustly outperforms the competing schemes \textcolor{\editagecolor}{for} \textcolor{\sryoonrevisioncolor}{most of the datasets tested.} We specifically emphasize that \acl~shows a strong performance\textcolor{\editagecolor}{, even though} the DMC assumption does not hold for the sequencer.
We believe that th\textcolor{\proofcolor}{e} \textcolor{\editagecolor}{better performance} of \acl~\textcolor{\editagecolor}{relative to other} state-of-the-art algorithms (\textcolor{\sryoonrevisioncolor}{based on MSA or $k$-mer spectrums}) on real Illumina datasets strongly demonstrates the competitiveness of \acl~as a general DNA sequence denoiser \textcolor{\sryoonrevisioncolor}{for targeted amplicon sequencing}.

\subsection*{Experiments on Simulated Data.}
\textcolor{\tsmcolor}{\textcolor{\editagecolor}{W}e \textcolor{\editagecolor}{performed} more detailed experiments using \textcolor{\tsmcolorfinal}{Illumina simulators} in order to further highlight the strong denoising performance of \acl, including the effect\textcolor{\editagecolor}{s} on downstream analyses.}

\begin{figure}[!h]
	\centering
	\includegraphics[width=0.5\linewidth]{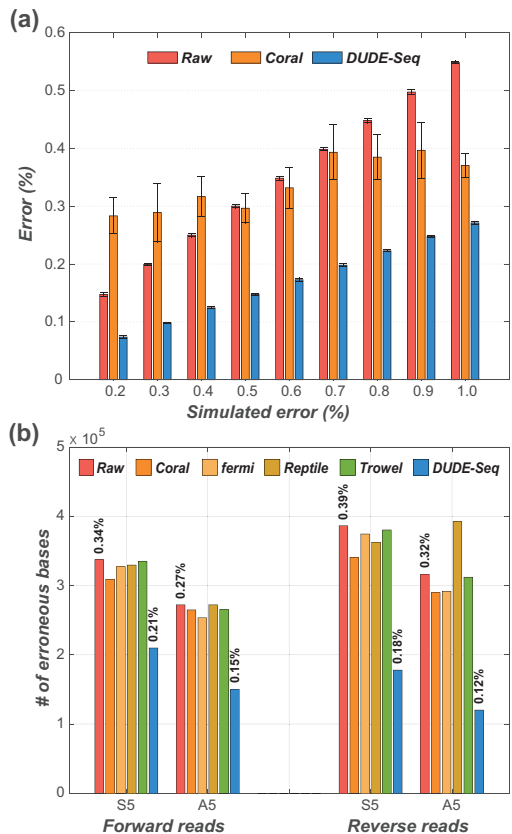}
	\begin{spacing}{\mylinespacing}
		\caption{{\bf Reads correction performance on simulated dataset.} {[parameters: $k=5$ for \acl; $k=10$ for Trowel; $(k,mr,mm,g)=(21,1,1,1000)$ for Coral; optimal values set by tool \texttt{seq-analy} for Reptile; \textcolor{\sryoonrevisioncolor}{$(k,O,C,s)=(21,3,0.3,5)$ for fermi}]}: (a) Varying error rates using the Grinder simulator~\cite{angly2012grinder}. (b) Varying reads composition using the GemSIM simulator~\cite{mcelroy2012gemsim} (values on top of each bar represent the error rates).}
		\label{fig:results-varying-error}
	\end{spacing}
\end{figure}

\textcolor{\tsmcolor}{Fig~\ref{fig:results-varying-error}(a) shows the results obtained using the Grinder simulator~\cite{angly2012grinder} and \textcolor{\editagecolor}{a} comparison
with Coral.
\textcolor{\proofcolor}{Trowel and Reptile require quality scores as input, which are provided by the GemSIM simulator, but not by the Grinder simulator; hence, we could not include Trowel and Reptile in Fig~\ref{fig:results-varying-error}(a).}
We generated nine synthetic datasets of forward reads that had error rates at the end of the sequence varying from 0.2\% to 1.0\%, as denoted \textcolor{\editagecolor}{on} the horizontal axis. For all cases, the error rate at the beginning of the sequence was 0.1\%. We again used the \emph{average} DMC model for the entire sequence for \acl. }
\textcolor{black}{Note that the error rates for the `Raw' data, \ie, the red bars, match the average of the error rates at the beginning and the end of the sequence.} \textcolor{\tsmcolor}{From the figure, \textcolor{\editagecolor}{consistent} with the real datasets \textcolor{\editagecolor}{analyzed} in Section \ref{sec:real_illumina}, we clearly see that \acl~significantly outperforms Coral \textcolor{\editagecolor}{for all} tested error rates.}

\textcolor{\editagecolor}{T}o evaluate the performance of \acl~for paired-end reads, we generated datasets\textcolor{\editagecolor}{, shown} in \tablename~\ref{tab:illumina-data}\textcolor{\editagecolor}{,} with the GemSIM sequencing data simulator~\cite{mcelroy2012gemsim}. As shown in the table, we used 23 public reference sequences~\cite{quince2011removing} \textcolor{\editagecolor}{to generate} the dataset A5 and a single reference sequence for S5. \textcolor{black}{We used the error model v5 that has error rate\textcolor{\editagecolor}{s} of 0.28\% for forward reads and 0.34\% for reverse reads.}
In Fig~\ref{fig:results-varying-error}(b), in addition to \acl, Coral, \textcolor{\tsmrevisioncolor}{fermi}, {and Trowel}, we included the result\textcolor{\editagecolor}{s obtained using} Reptile~\cite{yang2010reptile}, another $k$-mer spectrum\textcolor{\editagecolor}{-}based method that outputs read-by-read denoising results.
We again observe from the figure that \acl~outperforms \textcolor{\tsmrevisioncolor}{the alternatives} \textcolor{\proofcolor}{by} significant margins.

\setlength{\tabcolsep}{10pt}
\ctable[
	pos = !ht,
    caption = {\textcolor{\bhlcolor}{\bf Details of the public data \cite{kwon2014casper} used for our experiments on simulated data \textcolor{\srycolor}{shown in \tablename~\ref{tab:paired-end-merging-result}}}},
    label = tab:illumina-data,
    doinside = \scriptsize
]{crrcccc}{
\tnote[$\sharp$]{\small{Error model v5 (forward rate 0.28\%, reverse 0.34\%)}}
}{
    \toprule
    dataset & \# total & \multicolumn{1}{c}{\multirow{2}[0]{*}{\# refs}} & fragment & read  & overlap & simulator (error model) \\

    ID    &  reads  & \multicolumn{1}{c}{} & length & length & length & or sequencer used \\
    \midrule

    S5    &       1,000,000  & 1     & 160   & 100   & 40    & GemSIM (v5$^\sharp$) \\
    A5    &       1,000,000  & 23    &  160--190  & 100   & 10--40 & GemSIM (v5$^\sharp$) \\
    \bottomrule
    \vspace{-0.8em}
}

In \tablename~\ref{tab:paired-end-merging-result}, we show that the error-corrected reads produced by \acl~can also improve the \textcolor{\editagecolor}{performance of} downstream pipeline\textcolor{\editagecolor}{s}, such as paired-end merging. We applied four different paired-end merging schemes, CASPER~\cite{kwon2014casper}, COPE~\cite{liu2012cope}, FLASH~\cite{magovc2011flash}, and PANDAseq~\cite{masella2012pandaseq}, for the two datasets A5 and S5 in \tablename~\ref{tab:illumina-data}.
The metrics are defined as usual. \textcolor{\proofcolor}{A} true positive (TP) is defined as a merge with correct mismatching resolution in the overlap region, and a false positive (FP) is defined as a merge with incorrect mismatching resolution in the overlap region. Furthermore, a false negative (FN) is a merge that escapes the detection, and a true negative (TN) is defined as \textcolor{\editagecolor}{a} correct prediction \textcolor{\editagecolor}{for} reads that do not truly overlap. The accuracy and F1 score are computed based on the above metrics \cite{witten2005data}.
\textcolor{black}{For each dataset, we compared the results for \textcolor{\editagecolor}{four} cases: performing paired-end merging without any denoising, after correcting errors with Coral, after correcting errors with fermi, and after correcting errors with \acl.
Reptile and Trowel were not included in this experiment \textcolor{\proofcolor}{because} \textcolor{\editagecolor}{they were generally outperformed by} Coral and fermi\textcolor{\proofcolor}{,} as shown in Fig~\ref{fig:results-varying-error}(b).
The accuracy and F1 score results show that correcting errors with \acl~consistently yields better paired-end merging performance, not only compared to the \textcolor{\editagecolor}{case with} no denoising, but also \textcolor{\proofcolor}{compared} to the \textcolor{\editagecolor}{cases with} Coral and fermi \textcolor{\proofcolor}{error correct\textcolor{\editagecolor}{ion}}. This result highlights the potential \textcolor{\editagecolor}{application of} \acl~for boosting the performance \textcolor{\editagecolor}{of} downstream DNA sequence \textcolor{\editagecolor}{analyses}.}

\setlength{\tabcolsep}{8pt}
\ctable[
	pos= !ht,
    caption = {\textcolor{\bhlcolor}{\bf Paired-end reads merging performance statistics {[parameters: $k=5$ for \acl; $(k,mr,mm,g)=(21,1,1,1000)$ for Coral; $(k,O,C,s)=(21,3,0.3,5)$ for fermi]}}},
    label = {tab:paired-end-merging-result},
    doinside = \scriptsize,
]{lcrrrrrr}{
}{
    \toprule
    \multicolumn{1}{c}{tool} & \multicolumn{1}{c}{dataset} & \multicolumn{1}{c}{\# merges} & \multicolumn{1}{c}{TP} & \multicolumn{1}{c}{FP} & \multicolumn{1}{c}{FN} & \multicolumn{1}{c}{accuracy} & \multicolumn{1}{c}{$F_1$} \\
    \midrule

    \multicolumn{1}{l}{CASPER} & \multirow{4}[0]{*}{\shortstack{S5}}    &   1,000,000  &   997,303  &     2,697  &      \textbf{0}  &     0.997 &        0.999 \\
    \multicolumn{1}{l}{COPE}  &  &   974,219  &     961,366  &    12,853  &     25,781 &     0.961  &      0.980  \\
    \multicolumn{1}{l}{FLASH} &   &   999,921  &   977,431  &     22,490  &    79 &     0.977  &      0.989  \\
    \multicolumn{1}{l}{PANDAseq} &  &   999,947  &   976,807  &    23,140 &      53  &     0.977  &      0.988  \\
    \midrule

    \multicolumn{1}{l}{CASPER} & \multirow{4}[0]{*}{\shortstack{S5\\w/ Coral}}    &     1,000,000  &     997,510  &      2,490  &       \textbf{0}  &     0.998 &        0.999 \\
    \multicolumn{1}{l}{COPE}  &  &     975,803  &     963,717  &      12,086  &     24,197  &     0.964  &      0.982  \\
    \multicolumn{1}{l}{FLASH} &   &     999,942  &     978,835  &     21,107  &      58  &     0.979  &      0.989  \\
    \multicolumn{1}{l}{PANDAseq} &  &   999,949  &     978,270  &     21,679  &       51  &     0.978  &      0.989  \\
    \midrule

    \multicolumn{1}{l}{CASPER} & \multirow{4}[0]{*}{\shortstack{S5\\w/ fermi}}    &     1,000,000  &     997,356  &      2,644  &       \textbf{0}  &     0.997 &        0.999 \\
    \multicolumn{1}{l}{COPE}  &  &     994,025  &     969,451  &      24,574  &     \textbf{5,975}  &     0.969  &      0.984  \\
    \multicolumn{1}{l}{FLASH} &   &     999,933  &     972,025  &     27,908  &      67  &     0.972  &      0.986  \\
    \multicolumn{1}{l}{PANDAseq} &  &   999,952  &     971,567  &     28,385  &       48  &     0.972  &      0.986  \\
    \midrule

    \multicolumn{1}{l}{CASPER} & \multirow{4}[0]{*}{\shortstack{S5\\w/ \acl}}    &     1,000,000  &     \textbf{999,320}  &      \textbf{680}  &       \textbf{0}  &     \textbf{0.999} &        \textbf{1.000} \\
    \multicolumn{1}{l}{COPE}  &  &     987,238  &     \textbf{983,639}  &      \textbf{3,599}  &     12,762  &     \textbf{0.984}  &      \textbf{0.992}  \\
    \multicolumn{1}{l}{FLASH} &   &     999,958  &     \textbf{992,915}  &     \textbf{7,043}  &      \textbf{42}  &     \textbf{0.993}  &      \textbf{0.996}  \\
    \multicolumn{1}{l}{PANDAseq} &  &   999,949  &     \textbf{991,146}  &     \textbf{8,803}  &       \textbf{51}  &     \textbf{0.991}  &      \textbf{0.996}  \\
    \midrule

    \multicolumn{1}{l}{CASPER} & \multirow{4}[0]{*}{\shortstack{A5}}    &     999,973  &     997,202  &      2,771  &       27  &     0.997 &        \textbf{0.999} \\
    \multicolumn{1}{l}{COPE}  &  &     924,634  &     915,981  &      8,653  &     75,366  &     0.916  &      0.956  \\
    \multicolumn{1}{l}{FLASH} &   &     999,578  &     977,355  &     22,223  &      422  &     0.977  &      0.989  \\
    \multicolumn{1}{l}{PANDAseq} &  &   999,122  &     978,720  &     20,402  &       878  &     0.979  &      0.989  \\
    \midrule

    \multicolumn{1}{l}{CASPER} & \multirow{4}[0]{*}{\shortstack{A5\\w/ Coral}}    &     999,974  &     995,899  &      4,075  &       \textbf{26}  &     0.996 &        0.998 \\
    \multicolumn{1}{l}{COPE}  &  &     927,757  &     918,733  &      9,024  &     72,243  &     0.919  &      0.958  \\
    \multicolumn{1}{l}{FLASH} &   &     999,742  &     978,814  &     20,928  &      \textbf{258}  &     0.979  &      0.989  \\
    \multicolumn{1}{l}{PANDAseq} &  &   999,351  &     979,899  &     19,452  &       649  &     0.980  &      0.990  \\
    \midrule

    \multicolumn{1}{l}{CASPER} & \multirow{4}[0]{*}{\shortstack{A5\\w/ fermi}}    &     999,969  &     997,288  &      2,681  &       31  &     0.997 &        \textbf{0.999} \\
    \multicolumn{1}{l}{COPE}  &  &     939,986  &     923,252  &      16,734  &     60,014  &     0.923  &      0.960  \\
    \multicolumn{1}{l}{FLASH} &   &     999,732  &     974,903  &     24,829  &      268  &     0.975  &      0.987  \\
    \multicolumn{1}{l}{PANDAseq} &  &   999,328  &     975,320  &     24,008  &       672  &     0.975  &      0.988  \\
    \midrule

    \multicolumn{1}{l}{CASPER} & \multirow{4}[0]{*}{\shortstack{A5\\w/ \acl}}    &   999,971  &   \textbf{998,078}  &     \textbf{1,893}  &      29  &     \textbf{0.998} &        \textbf{0.999} \\
    \multicolumn{1}{l}{COPE}  &  &   943,531  &     \textbf{939,555}  &    \textbf{3,976}  &     \textbf{56,469} &     \textbf{0.940}  &      \textbf{0.969}  \\
    \multicolumn{1}{l}{FLASH} &   &   999,638  &   \textbf{989,860}  &     \textbf{9,778}  &    362 &     \textbf{0.990}  &      \textbf{0.995}  \\
    \multicolumn{1}{l}{PANDAseq} &  &   999,354  &   \textbf{989,250}  &    \textbf{10,104} &      \textbf{646}  &     \textbf{0.989}  &      \textbf{0.995}  \\
    \bottomrule
}

\section*{Discussion}
\textcolor{black}{
Our experimental results show that \acl~can robustly outperform $k$-mer\textcolor{\editagecolor}{-}based, MSA\textcolor{\proofcolor}{-}based, and statistical error model\textcolor{\editagecolor}{-}based schemes \textcolor{\editagecolor}{for} both real-world datasets, such as 454 pyrosequencing and Illumina \textcolor{\editagecolor}{data}, and simulated datasets, \textcolor{\tsmrevisioncolor}{particularly for targeted amplicon sequencing.}
This performance advantage in denoising further allowed us to obtain improved results in downstream analysis tasks, such as OTU binning and paired-end merging. Furthermore, the time demand of \acl-based OTU binning is order(s) of magnitude lower than that of the current state-of-the-art \textcolor{\proofcolor}{schemes}. We also demonstrated \textcolor{\srycolorfinal}{the robustness and flexibility of \acl~by showing} that a simple change \textcolor{\proofcolor}{in} $\mathbf{\Pi}$ matrix is enough to apply the exact same \acl~to data \textcolor{\editagecolor}{obtained using} different sequencing platforms.} \textcolor{black}{In particular, we experimentally showed that even when the memoryless channel assumption does not hold, as in Illumina \textcolor{\editagecolor}{data}, \acl~still solidly outperforms state-of-the-art schemes.}

\textcolor{black}{The sliding window nature of \acl~resemble\textcolor{\editagecolor}{s} the popular $k$-mer\textcolor{\editagecolor}{-}based schemes in the literature. However, while all existing $k$-mer\textcolor{\editagecolor}{-}based schemes rely on heuristic threshold \textcolor{\editagecolor}{selection} for determining errors in the reads\textcolor{\proofcolor}{,} regardless of the error model of the sequencing platform, \acl~applies an analytic denoising rule that explicitly takes the error model $\mathbf{\Pi}$ into account. Therefore, even for \textcolor{\editagecolor}{identical} noisy reads $z^n$, DUDE-Seq may result in different denoised sequences\textcolor{\proofcolor}{,} depending on the $\mathbf{\Pi}$'s of different sequencing platforms, whereas the $k$-mer\textcolor{\editagecolor}{-}based scheme will always result in the exact same denoised sequence.
The performance gains reported in this paper compared to state-of-the-art baselines, including \textcolor{\proofcolor}{those for} $k$-mer\textcolor{\editagecolor}{-}based schemes, substantiate the competitiveness of our method for \textcolor{\tsmrevisioncolor}{targeted amplicon sequencing.}}

\textcolor{black}{
Another advantage of \acl~is its read-by-read error-correction capability. This feature is important for a number of bioinformatics tasks\textcolor{\proofcolor}{,} including \emph{de novo} sequencing, metagenomics, resequencing, targeted resequencing, and transcriptome sequencing, which typically require \textcolor{\editagecolor}{the} extract\textcolor{\editagecolor}{ion of} subtle information from small variants in each read. In addition to the types of tasks presented in this paper (\textcolor{\proofcolor}{\ie,} per-based error correction, OTU binning, and paired-end merging), we plan to apply \acl~to additional tasks\textcolor{\proofcolor}{,} as mentioned above.}

Additional venues for further investigation include the procedure for estimating the noise mechanism represented by $\mathbf{\Pi}$, which is currently empirically determined by aligning each read to the reference sequence and is \textcolor{\proofcolor}{therefore} sensitive to read mapping and alignment. For more robust estimation, we may employ an expectation-maximization-based algorithm, as was recently proposed for estimating substitution emissions for the data \textcolor{\editagecolor}{obtained using} nanopore technology~\cite{jain2015improved}.
Considering uncertainties in $\mathbf{\Pi}$ may also be helpful; hence, it may be useful to investigate the relevance of the framework in \cite{gemelos2006algorithms}.
Additionally, it will likely be fruitful to utilize the information \textcolor{\editagecolor}{in} Phred quality scores \textcolor{\editagecolor}{to make} decisions about noisy bases and \textcolor{\editagecolor}{to} fine-tun\textcolor{\editagecolor}{e} the objective loss function in our approach.
\textcolor{black}{Using a lossy compressed version of the quality scores \textcolor{\editagecolor}{is} one possible direction \textcolor{\proofcolor}{for} boost\textcolor{\proofcolor}{ing} the inferential performance of some downstream applications\textcolor{\editagecolor}{,} as shown in~\cite{ochoa2015effect}.}
\textcolor{black}{Furthermore, particularly for the homopolymer error correction, there are several hyperparameters whose choices can be experimented with in the future \textcolor{\proofcolor}{to} potentially \textcolor{\proofcolor}{achieve} substantial performance boosts. \textcolor{\proofcolor}{Examples include} the choice of alphabet size (in lieu of the current value of 10), the choice of the loss function that may be proportional to the difference between the true and estimated value of $N$ (in lieu of the current Hamming loss), and the choice of quantization (in lieu of (\ref{eq:rounding})).
Moreover, we may apply the full generalized DUDE in \cite{dembo2005universal} for homopolymer error correction to \textcolor{\editagecolor}{determine} if better performance can be achieved at the cost of increased complexity.}
\textcolor{\bhlploscolor}{Applying \acl~to other types of sequencing technology with homopolymer errors (\eg, Ion Torrent) would also be possible \textcolor{\editageploscolor}{as} long as we can acquire \textcolor{\bhlplosrevisioncolor}{flow (\eg, ionogram)} density distributions to estimate $\mathbf{\Gamma}$.
Currently, there exists no public data repository that \textcolor{\editageploscolor}{includes} such information for Ion Torrent, and thus existing Ion Torrent denoisers often ignore homopolymer errors or rely on simplistic noise modeling and iterative updates that unrealistically limit the maximum length of homopolymer errors that can be handled, let alone computational efficiency~\cite{bragg2013shining}.
}
\textcolor{\tsmcolor}{Finally, we plan to test \acl~on several other sequencing platforms\textcolor{\proofcolor}{,} such as PacBio and Oxford Nanopore, which tend to result in longer and more noisy sequences, to further substantiate the robustness and effectiveness of our algorithm. \textcolor{\tsmrevisioncolor}{Applying the recently developed deep neural networks\textcolor{\proofcolor}{-}based Neural DUDE algorithm \cite{MooMinLeeYoo16} to DNA sequence denoising beyond targeted amplicon sequencing could be another fruitful direction.}}

\section*{Supporting Information}

\paragraph*{S1 Fig.}\label{S1_Fig}
\textcolor{\srycolor}{
{\bf \acl~web interface.}
This is a screenshot of the website accompanying the proposed \acl~method ( \href{http://data.snu.ac.kr/pub/dude-seq}{http://data.snu.ac.kr/pub/dude-seq}). For users who prefer a graphical user interface, this website provides a web-based execution environments for \acl. Through this screen, a user can specify the parameters for each of the two error types (in the figure, \acl~(1) stands for for the substitution error correction described in Algorithm~1 and \acl~(2) stands for the homopolymer error correction shown in Algorithm~2), and upload the input file of her choice. The \acl~process starts automatically by clicking the ``SUBMIT'' button. For advanced users who prefer batch processing, the source code of \acl~is also available at \href{http://github.com/datasnu/dude-seq}{http://github.com/datasnu/dude-seq}.
}

\paragraph*{S2 Fig.}\label{S2_Fig}
\textcolor{\srycolor}{
{\bf Website output: sequence complexity.}
The \acl~website provides analysis results from applying the DUST algorithm~\cite{morgulis2006fast} and block-entropy to the outputs from denoising by \acl. The DUST algorithm masks low-complexity regions that have highly biased distribution of nucleotides based on counting 3-mer frequencies in 64-base windows. \textcolor{\bhlplosrevisioncolor}{The DUST score is computed based on how often different trinucleotides occur as follows:
\begin{equation}
    \text{score} = \sum_{i=1}^{k} \frac{n_i(n_i-1)(w-2)s}{2(l-1)l} \nonumber
\end{equation}
where $k=4^3$ is the trinucleotide size, $w=64$ is the window size, $n_i$ is the number of the words $i$ in a window, $l$ is the number of the possible words in a window, and $s$ is the scaling factor. The score is scaled from 0 to 100 and a high score implies a low complexity metagenome.}
The block-entropy is calculated using Shannon's diversity index~\cite{shannon2001mathematical}. \textcolor{\bhlplosrevisioncolor}{The block-entropy evaluates the entropy of the trinucleotides in a sequence as follows:
\begin{equation}
    \text{entropy} = -\sum_{i=1}^{k} \big(\frac{n_i}{l}\big) \text{log}_k \big(\frac{n_i}{l}\big) \nonumber
\end{equation}
where $k=4^3$ is the trinucleotide size, $n_i$ is the number of the words $i$ in a window, and $l$ is the number of the possible words in a window. The entropy is also scaled from 0 to 100 and a low entropy implies a low complexity metagenome.}
}

\paragraph*{S3 Fig.}\label{S3_Fig}
\textcolor{\srycolor}{
{\bf Website output: tag sequence probability.}
Another output from the \acl~website is the tag sequence probability of reads~\cite{schmieder2010tagcleaner}. This is to reveal the existence of artifacts at the ends, \ie, adapter or barcode sequences at the 5'- or 3'-end.
}

\paragraph*{S4 Fig.}\label{S4_Fig}
\textcolor{\srycolor}{
{\bf Website output: sequence duplication.} The accompanying website also carries out sequence duplication analysis based on the denoised outputs from \acl, in order to reveal artificial duplicates. As shown in the figure, five types of duplication statistics~\cite{schmieder2011quality} are provided: exact duplicates, 5' duplicates, 3' duplicates, exact duplicates with the reverse complement of another sequence, and 5' or 3' duplicates with the reverse complement of another sequence.
}

\section*{Acknowledgments}
This work was supported in part by the National Research Foundation of Korea (NRF) grant funded by the Korea government (Ministry of Science, ICT and Future Planning) [No. 2014M3A9E2064434], in part by a grant of the Korea Health Technology R\&D Project through the Korea Health Industry Development Institute (KHIDI), funded by the Ministry of Health \& Welfare, Republic of Korea [HI14C3405030014 and HI15C3224], and in part by the Brain Korea 21 Plus Project in 2017.

\nolinenumbers

\newcommand{\noop}[1]{}

\end{document}